\documentclass[aps,eqsecnum,groupedaddress,a4paper,floatfix,twocolumn,footinbib]{revtex4}
\usepackage{amsfonts}
\usepackage{amssymb}
\usepackage{mathrsfs}
\usepackage[mathscr]{euscript}
\usepackage[dvips]{graphicx}
\usepackage{fancyhdr}
\begin{document}
\title{Quantum stochastic theory of phonon scattering between polaritons}
\author{P. Kinsler}

\altaffiliation[This research was primarily done at the ]{
Department of Physics, University of Sheffield, Sheffield S3 7RH,
United Kingdom in 1996.  I have 
released now it as an eprint because although it could be made publishable 
in a refereed journal, the work has been dormant for a few years and I can 
no longer find the time to do the necessary work 
}
\email{Dr.Paul.Kinsler@physics.org}

\affiliation{
Department of Physics, Imperial College,
  Prince Consort Road,
  London SW7 2BW, 
  United Kingdom}

\lhead{EPRINT}
\chead{~}
\rhead{Dr.Paul.Kinsler@physics.org\\
http://www.kinsler.org/physics/}

\date{\today}

\begin{abstract}

  Quantum stochastic operator equations are derived for inter-branch exciton
and polariton processes caused by acoustic phonon scattering.  The use of a
fully quantum model combined with these recently developed techniques predicts
the presence of ``stimulated scattering'' terms, and provides a sound basis for
understanding the basis of the approximations used in generating the equations.
The theory is applied to a model motivated by recent experiments where a
stronger photoluminescence signal from the high energy microcavity polariton
polariton is observed for low excitation powers, and also to an experimental
scheme designed to show evidence of the stimulated scattering effect.

\end{abstract}

\maketitle
\thispagestyle{fancy}


\begin{section}{Introduction}


  The  development of stochastic operator and wavefunction approaches to
dissipative processes caused much interest in the field of quantum
optics.  My aim here is to present a derivation of equations for inter-branch
phonon processes using one of these techniques, that of quantum stochastic
differential equations (QSDE's) \cite{Gardiner-qn}.  The derivation
of these equations relies on similar approximations to the well known master
equation techniques \cite{Gardiner-qn,Louisell-1973}, and so
has the advantage of being a systematic derivation in which the approximations
are clearly stated and whose effects are generally well understood.  Master
equation techniques from a ``quantum optical'' viewpoint have already been
applied to both excitonic optical bistability \cite{Steynross-G-1983} and the
semiconductor laser \cite{Gardiner-E-1995}.  The stochastic wavefunction
approach based on an ``unravelling'' of the master equation
\cite{Carmichael-1993} has been applied to both electron-phonon scattering
\cite{Inamoglu-Y-1994} and exciton-phonon scattering \cite{Stenius-I-1996}. 
This exciton-phonon work was for a model involving a single quantum-dot exciton
level, and the emphasis was on non-Markovian effects.  

  Here the QSDE technique is used to treat the phonon interaction as a coupling
between two exciton or polariton branches and an infinite reservoir of phonon
modes.  Polaritons are formed in a semiconductor quantum microcavity,
structures in which quantum wells are grown inside an optical cavity.  The
light field is confined by two Bragg mirrors, and the quantum wells create
confined exciton states.  With appropriate design, the photons in
the optical cavity and the excitons couple together creating a quasi
particle called a cavity polariton.  The QSDE's for the polariton system
include factors accounting for its partly photonic nature.  In addition, they
contain extra ``stimulated scattering'' terms, also predicted by Pau et al 
\cite{Pau-BJCY-1995}, whose presence would not have been
predicted by a phemonenonological approach to modelling the system.

 This paper was motivated by photoluminescence (PL)
measurements of microcavity polaritons at low temperatures (1.7 -- 50K) and a
range of input laser intensity (0.4 -- 10 Wcm$^{-2}$) \cite{Tribe-etal}. 
Most investigations of these cavity polaritons use reflectivity 
measurements, with only a few considering the other optical emission 
properties.  Contrary to expectations, it was the {\em higher} energy 
polariton that dominated the PL at the lowest laser power densities.  
The experimental results are due to the thermalisation of the free 
carriers down into the upper polariton branch and the dominance of the 
cavity decay rate over the scattering between the polariton branches.  
Free carrier scattering becomes insignificant at low carrier densities
and LO phonon processes do not
occur because their energies are greater than the polariton splitting.  
As a result, the photon intensity (PL signal) emitted from decay 
of the upper polariton is strong, while the signal from the 
lower polariton is restricted by the small influx of
excitation due to the slow inter-branch acoustic phonon scattering.  
The equations for the
theoretical model are solved in an appropriate limit to predict the severity of
this bottleneck.  The theory is also extended to a system where the upper
polariton is coherently excited by a laser tuned to the polariton energy, and
predicts a similar bottleneck effect.   The results depend on the size of the
photonic component of the polariton as well as its excitonic part.  

 Finally, the stimulated scattering term is investigated by constructing a
model experimental system that could be used to measure its effects.  This
model also predicts the strength of the effect, which is proportional to the
exciton- phonon coupling, the exciton- photon mixing, and the intensity of the
stimulating field.

~

\noindent
{\bf Note added (March 2002)}

Stimulated emission of bulk polaritons was experimentally observed in II-VI
sample in the late 1970's \cite{Klingshirn-H-1981pr}, and microcavity
polaritons in QW samples in 1997 \cite{Kelkar-KNCHG-1997prb}.  Stimulated
emission terms (i.e. a factor of $N+1$) were inserted by hand into the theory
of Tassone et. al. \cite{Tassone-PSQS-1997prb} on the bottleneck effects in
microcavity polaritons, and then with more theoretical justification in by
Tassone and Yamamoto \cite{Tassone-Y-1999prb} which included discussion of
stimulated scattering.  In these two papers the interest is restricted to
generating a rate equation model, and the papers pay little theoretical
attention to all the steps in a rigorous derivation such as I have attempted
here, and certainly quantum noise terms are not considered at all.

The bottleneck effects and stimulated scattering have become of much recent
interest -- see for example some of the references brought to my attention by
J. J. Baumberg
\cite{Baumberg-ASR-199prl,Savvidis-BSSWR-2000prb,Tartakovskii-ESSAWBR-2000prb}
.  It turns out the model I used when doing these early calculations at
University of Sheffield in 1996 is far too simplistic, and so is not relevant
to any likely experimental work.  However, the theoretical method used to
generate both the QSDE's and model is still valid, and it would be a 
straightforward matter to recast the model to a more appropriate one and 
do some useful calculations.

\end{section}

\begin{section}{The phonon interaction}


 In this section I introduce a model Hamiltonian for the phonon scattering
process.  This does not specify the type of phonon explicitly, they could be
either optical or acoustic (either longitudinal or transverse), as the 
phonons are treated as a reservoir of boson
modes with particular properties, properties that are necessary for the QSDE
technique to be valid.  As long as the type of phonons considered have these
properties, the results given here are applicable.

 It is also relevant to consider here how we describe the exciton and cavity
modes.  For low excitation strengths the exciton can be described as
a boson.  This is because the resulting low density of excitons means that any
phase space filling \cite{Haug-S-1984} or exciton-exciton scattering processes
are unlikely to occur.  
In addition, there is a continuum of exciton modes for different
in-plane momenta.  I will refer to this as an 'exciton branch', and the
energies of these are given by the dispersion relations for the exciton. 
Similarly, a microcavity photon mode can also have a range of $k$ values, and
can also be called a 'photon branch'.

 Consider an interaction Hamiltonian for quantum well excitons interacting with
phonons.  The two dimensional nature of each exciton branch is allowed for by a
set of exciton modes indexed by the in-plane wave vector $k$, an these modes 
are associated with the operators $\hat{e}_i (k)$, $\hat{e}_i ^\dagger (k)$.  
The phonon modes are indexed by in-plane wave vector $k$ as well as
their growth direction wave vector $q$, with the operators $\hat{b}(k, q)$, 
$\hat{b} ^\dagger (k, q)$. 
The coupling between exciton and phonons is denoted $\chi$.  In the interaction
picture, the operators all have an oscillatory time dependence given by their
mode energies, and so $\hbar \omega \hat{a} ^\dagger \hat{a}$ type terms do not
appear in the Hamiltonian.  This time dependence is not explicitly included to
avoid cluttering the notation.  The general form of the interaction Hamiltonian
for first order scattering involving any two exciton branches and one phonon 
branch is


\begin{eqnarray}
\hat{H}_A 
&=&
\sum_{ij}
\sum_{k} \sum_{k'} \sum_{k''} \sum_{q}
   \chi_{ij} (k, k', k'', q)
\nonumber
\\
&&   \left[
      \hat{e}_i (k) + \hat{e}_i ^\dagger (k) 
   \right]
   \left[
      \hat{e}_j ^\dagger (k') + \hat{e}_j (k') 
   \right]
\nonumber
\\
&&   \left[
      \hat{b} ^\dagger (k'', q) + \hat{b} (k'', q)
   \right]
\end{eqnarray}

 Often the terms are written like $\left[ \hat{e}_i (k) + 
\hat{e}_i ^\dagger (-k) \right]$, but this is simply a re-ordering 
of the same summation.  The $i$, $j$ indexes span the exciton branches, 
the $k$ summations are
over the in-plane wave vector of the excitons, and the $q$ sum is over the wave
vector of the phonon in the growth direction.  Note that with only minor
modifications to the summations and indices the same expression can be used to
describe phonon scattering in bulk or even 1 or 0 dimensional structures.  

 The first simplification we can make is to apply conservation of in-plane
crystal momentum $k$, which is not conserved because of the lack 
of translational symmetry in the growth direction.  This reduces the 
number of summations in the above formula from three to two, and results in 

\begin{eqnarray}
\hat{H}_A
&=&
\sum_{ij}
\sum_{k} \sum_{\Delta k} \sum_{q}
   \chi_{ij} (k, k-\Delta k, q)
\nonumber
\\
&&   \left[
      \hat{e}_i (k) \hat{e}_j ^\dagger (k - \Delta k) 
      \hat{b} ^\dagger (\Delta k, q)
\right.
\nonumber
\\
&&\left.
      +
      \hat{e}_i ^\dagger (k) \hat{e}_j (k - \Delta k) 
      \hat{b} (\Delta k, q)
   \right.
\nonumber \\
& &
   \left.
      +
      \hat{e}_i ^\dagger (-k) \hat{e}_j ^\dagger (k - \Delta k) 
      \hat{b} ^\dagger (\Delta k, q)
\right.
\nonumber
\\
&&\left.
      +
      \hat{e}_i (-k) \hat{e}_j (k - \Delta k) 
      \hat{b} (\Delta k, q)
   \right]
.
\end{eqnarray}

\chead{Quantum stochastic theory of phonon scatt...}

 It is normal in quantum optics to make a rotating wave approximation to 
remove non-resonant (non energy conserving) terms because these
off resonant terms oscillate at about twice the optical frequency.  
This is very much
faster than any likely dynamical process in the system, and so their net effect
is assumed to average to zero, in a kind of ``coarse graining'' approximation. 
In this system involving phonons, it is not obvious that this type of
approximation can be made.  Both the excitons and the phonons have a
continuum of energies, and because the phonons can have frequencies near zero,
the oscillations in the off resonant terms can easily be relatively slow, so
the rotating wave approximation (RWA) cannot always be applied.  A situation
like this can be dealt with using non-rotating wave techniques (eg. 
\cite{Munro-G-1996})

 In the case of optical phonon scattering a RWA {\em is} possible.  This is
because optical phonons have a minimum frequency (about $10^{13}$ Hz), and the
off resonant terms would be oscillating at about twice that rate.  Also,
acoustic phonon scattering could be treated between two sufficiently separated
exciton branches, as again the off resonant terms would oscillate at twice the
frequency of the separation of the two excitons.  Alternatively, if for some
reason the exciton to phonon coupling vanished sufficiently fast for small
phonon energies the exciton could only couple to relatively energetic phonons
for which a RWA might be reasonable.  Assuming well separated branches 
or optical
phonons, the oscillations in the off resonant terms will occur on much shorter
timescales than any other dynamical process in the system.  This means we can
apply the RWA, so the Hamiltonian becomes

\begin{eqnarray}
\hat{H}_A
&=&
\sum_{i \neq j}
\sum_{k} \sum_{\Delta k} \sum_{q}
   \chi_{ij} (k, k-\Delta k, q)
\nonumber
\\
&&
   \left[
      \hat{e}_i (k) \hat{e}_j ^\dagger (k - \Delta k) 
      \hat{b} ^\dagger (\Delta k, q)
\right.
\nonumber
\\
&&\left.
      +
      \hat{e}_i ^\dagger (k) \hat{e}_j (k - \Delta k) 
      \hat{b} (\Delta k, q)
   \right]
.
\end{eqnarray}


\begin{subsection}{The phonon QSDE}

 Following Gardiner \cite{Gardiner-qn}, we can write down the Ito quantum
stochastic differential equations (QSDE's) for the above model. In this
approach we treat the phonon modes as a reservoir or 'heat bath' with which the
excitons can interact.  At first sight, this QSDE method does not seem directly
applicable to this problem.  This is because there is only one reservoir of
phonons in the system, whereas the standard application of the theory would
assume that each scattering between excitons of different wave number is
coupled to an independent reservoir of phonons.  For this reason some
correlations might persist (or perhaps form) between different exciton modes
because they interact with this common reservoir.  We might reasonably 
assume this to be an insignificant effect because the phonon reservoir
is much bigger than the range of exciton states -- the phonon reservoir is
three dimensional, whereas the exciton states only fill two dimensions.  

 For example, compare this with the standard quantum optics case, where 
a single system mode interacts with a one dimensional continuum 
of reservoir modes -- there are an infinite number of reservoir modes 
for the system mode to decay into.  Similarly, the exciton
phonon scattering model presented here also has sufficiently large reservoirs
in which to ``lose'' any quantum correlations.  Although plausible enough, this
argument is not sufficient.  Leaving the detailed calculations to a later work
here I just make the extra assumption that any such
correlations will decay away, and so the use of QSDE's here is valid.

 There are three properties this reservoir must have for the QSDE approach to
be valid.  Firstly, there must be a smooth, dense spectrum of phonons with
which each exciton mode can interact.  For acoustic phonons, which have an
energy proportional to the magnitude of their wave vector, such a spectrum
exists. For LO phonons, the phonon modes only exist above a certain energy.
This can cause problems due to the effect of the edge of the LO phonon
band, but if the inter-exciton transition is sufficiently larger than the 
lowest LO phonon energy, the scattering will not see the phonon band edge.

 Secondly, the coupling between system and reservoir must be linear in the
reservoir operators, in agreement with the first order interaction
Hamiltonian used above.  

 Thirdly, the coupling constants between the reservoir and system must be
smooth functions of frequency, which is holds for deformation potential
coupling to phonons \cite{Mahan-1990}.  This leads to the additional and
related approximation usually called the First Markov Approximation.  This
requires the coupling parameter in the Hamiltonian between the excitons and the
reservoir of phonons to be independent of frequency in the range of interest. 
This ``range of interest'' is where energy is very nearly conserved -- where the
phonon carries away (or supplies) the energy lost (or gained) by the conversion
of the exciton.  Note that the coupling is assumed to remain roughly constant
over the width of the resonance. This is also a kind of ``coarse graining'', 
as a flat frequency response implies a strongly peaked time response -- 
so the coupling appears ``flat'' if the width of the time response is much 
shorter than the important dynamical processes occuring in the system. 
Also, the state of the reservoir must be
unaffected by the interaction.  In this case, the energy lost by the excitons
in the form of phonons must not change the state of the reservoir -- for
example, by increasing its temperature.  This is guaranteed for low excitation
levels by the relative sizes of the system and reservoir. 

  The procedure has allowed us to eliminate the phonon modes from the model. 
It has given us a system in which the phonon scattering causes a loss of energy
from the excitons and added a delta correlated ``quantum noise''.  This noise is
the effect of the quantum uncertainty of the phonon reservoir feeding into the
exciton system, and is described by new quantum noise operators $d
\hat{B}_P$, $d \hat{B}_P ^\dagger $.  


The Ito QSDE for some arbitrary system operator $\hat{S}$ affected by 
phonon scattering between two different exciton branches is

\begin{eqnarray}
&&d \hat{S}
=
\sum_{i \neq j} \sum_{\Delta k}
   \frac{ \chi _{ij} (k, k-\Delta k) }{2}
\nonumber \\
& &
\left\{ 
   \left[ N_P(\Delta k) + 1 \right]
   \left[
      2 
      \hat{e}_i ^\dagger (k) \hat{e}_j (k - \Delta k)
      \hat{S}
      \hat{e}_i (k) \hat{e}_j ^\dagger (k - \Delta k)
\right. \right.
\nonumber \\
& &
\left. 
   \left.
     -
      \hat{S}
      \hat{e}_i ^\dagger (k) \hat{e}_j (k - \Delta k) 
      \hat{e}_i (k) \hat{e}_j ^\dagger (k - \Delta k) 
   \right.
\right.
\nonumber \\
& &
\left.
   \left.
     -
      \hat{e}_i ^\dagger (k) \hat{e}_j (k - \Delta k) 
      \hat{e}_i (k) \hat{e}_j ^\dagger (k - \Delta k) 
      \hat{S}
   \right] 
\right.
\nonumber \\
& &
\left.
+
   N_P(\Delta k)
   \left[
      2 
      \hat{e}_i (k) \hat{e}_j ^\dagger (k - \Delta k)
      \hat{S}
      \hat{e}_i ^\dagger (k) \hat{e}_j (k - \Delta k)
   \right.
\right.
\nonumber \\
& &
\left.
   \left.
     -
      \hat{S}
      \hat{e}_i (k) \hat{e}_j ^\dagger (k - \Delta k) 
      \hat{e}_i ^\dagger (k) \hat{e}_j (k - \Delta k) 
   \right.
\right.
\nonumber \\
& &
\left.
   \left.
     -
      \hat{e}_i (k) \hat{e}_j ^\dagger (k - \Delta k) 
      \hat{e}_i ^\dagger (k) \hat{e}_j (k - \Delta k) 
      \hat{S}
   \right] 
\right\} dt
\nonumber \\
& &
 -
  \sum_{i,j} \sum_{\Delta k}
   \sqrt{ \chi _{ij} (k, k-\Delta k) }
   \left[ \hat{S}, \hat{e}_i ^\dagger (k) \hat{e}_j (k - \Delta k) \right]
   d \hat{B}_P
\nonumber \\
& &
 +
  \sum_{i,j} \sum_{\Delta k}
   \sqrt{ \chi _{ij} (k, k-\Delta k) }
   d \hat{B}_P ^\dagger
   \left[ \hat{S}, \hat{e}_i (k) \hat{e}_j ^\dagger (k - \Delta k) \right]
\nonumber \\
.
\end{eqnarray}

  This operator $\hat{S}$ could be any operator of interest -- perhaps one of
the mode operators, the intensity operator or even some kind of correlation
function.  The operators $d \hat{B}_P$, $d \hat{B}_P ^\dagger $ are quantum
white noise operators, and $N_P(\Delta k)$ is the average number of thermal
phonons in each phonon mode.  The summation over $q$ has dropped out because
the resonance condition allows us to calculate it in terms of the $k$ terms and
the dispersion curves.  The noise operators $d\hat{B}_P$, $d\hat{B}_P ^\dagger$
are defined by their correlations, which are

\begin{eqnarray}
\left< d\hat{B}_P (t) d\hat{B}_P (t') \right> 
&=& 
  \left< 
    \left[ d\hat{B}_P (t) ^\dagger d\hat{B}_P (t') ^\dagger 
    \right] 
  \right> = 0
,
\nonumber \\
\left<
  d\hat{B}_P (t) d\hat{B}_P ^\dagger (t') 
\right>
&=& 
  \left( N_P + 1 \right) \delta(t-t') dt
,
\nonumber \\
\left<
  d\hat{B}_P ^\dagger (t) d \hat{B}_P (t') 
\right>
&=& 
  N_P \delta(t-t') dt
.
\end{eqnarray}

  These operators cause ``quantum white noise'', and correspond to a reservoir
in which the number of thermal quanta is constant per unit bandwidth.  This is
not the same as a quantum thermal noise, which would have a Boltzmann
distribution for its thermal quanta.  Although using quantum white noise is not
exact, it is a very convenient and widely used approximation.  In particular,
it would not be a good approximation for the low wavevector part of an acoustic
phonon reservoir, where the large relative change in energy between nearby
wavevectors means a proper thermal ensemble should be used \cite{Gardiner-qn}. 
However, since we have already introduced a rotating wave approximation bases
on a large energy separation between exciton branches, and hence are not in the
low energy part of the phonon spectrum, using quantum white noise does not 
cause any extra restriction.

\end{subsection}

\begin{subsection}{Intensity and amplitude operator equations}

 To get a clearer picture of what the phonon interaction does, I will give some
examples relevant to the calculations in this paper.  In the quasi-PL and
coherent excitation models presented in section VI, equations for excitation
number and mode amplitude are used; more specifically, these are derived from
the QSDE's for the mode intensity and mode amplitude operators.  For
simplicity, I write $\hat{e}_i (k)$ as $\hat{e}_{+}$, $\hat{e}_j (k-\Delta k)$
as $\hat{e}_{-}$, and $\chi _{ij} (k, k-\Delta k)$ as $\chi$.  The intensity
for the $({+})$ mode is just $\hat{I}_{+} = \hat{e}_{+} ^\dagger \hat{e}_{+}$,
that for the $({-})$ mode is $I_{-} = \hat{e}_{-} ^\dagger \hat{e}_{-}$.  The
modes described by $\hat{e}_{+}$ and $\hat{e}_{-}$ are different.

The QSDE equations for the intensities of the $({+})$ and $({-})$ modes due to
the phonon scattering are

\begin{eqnarray}
d \hat{I}_{+}
&=&
\chi 
\left[ 
-  \hat{I}_{+} \left( \hat{I}_{-} + 1 \right) 
-  N_P
   \left( \hat{I}_{+} - \hat{I}_{-} \right)
\right] dt
\nonumber \\
&&
- \sqrt{\chi} 
\left[
   \hat{e}_{+} ^\dagger \hat{e}_{-} d \hat{B}_P 
 + d \hat{B}_P ^\dagger \hat{e}_{+} \hat{e}_{-} ^\dagger
\right]
,
\nonumber \\
d \hat{I}_{-}
&=&
\chi 
\left[ 
   \hat{I}_{+} \left( \hat{I}_{-} + 1 \right) 
+  N_P
   \left( \hat{I}_{+} - \hat{I}_{-} \right)
\right] dt 
\nonumber \\
&&
+ \sqrt{\chi} 
\left[
   \hat{e}_{+} ^\dagger \hat{e}_{-} d\hat{B}_P
 + d\hat{B}_P ^\dagger \hat{e}_{+} \hat{e}_{-} ^\dagger
\right]
.
\end{eqnarray}

  The noise terms cannot be rewritten in a form depending solely on
$\hat{I}_{\pm}$, so I leave them in terms of $\hat{e}_{+}$, $\hat{e}_{-}$. 
However, we could define new operators $\hat{J}_1 = \hat{e}_{+} ^\dagger 
\hat{e}_{-}$ and $\hat{J}_2 = \hat{e}_{+} \hat{e}_{-} ^\dagger$, and from them
we get extra equations which allow us to remove the $\hat{e}$ type terms.  
Note how the terms in the pair of equations balance each other to preserve
excitation number, a consequence of the fact that the scattering transfers
excitation between exciton branches while emitting (or absorbing) phonons. 
This contrasts with the usual loss models that use reservoirs which destroy
excitations by removing them from the system.  

  The QSDE equations for the amplitudes of the $({+})$ and $({-})$ modes due to
the phonon scattering are

\begin{eqnarray}
d \hat{e}_{+}
&=&
\chi 
\left[ 
  - \hat{e}_{+} \left( \hat{e}_{-} ^\dagger \hat{e}_{-} + 1 \right)
  - N_P \hat{e}_{+} 
\right] dt
+ \sqrt{\chi} \hat{e}_{-} d \hat{B}_P
,
\nonumber \\ 
d \hat{e}_{-}
&=&
\chi 
\left[ 
   \hat{e}_{+} ^\dagger \hat{e}_{+} \hat{e}_{-}
-  N_P \hat{e}_{-}
\right] dt 
+ \sqrt{\chi} d\hat{B}_P ^\dagger \hat{e}_{+}
.
\label{e-phonon-qnoise}
\end{eqnarray}

 These equations have an interesting feature -- the presence of a ``stimulated
scattering'' or ``incoherent stimulated emission'' term.  These show up as the
$\hat{I}_{+} \hat{I}_{-}$ and $N_P \left( \hat{I}_{-} - \hat{I}_{+} \right)$ in
the intensity equations.  A naive consideration of the phonon scattering
process would not suggest such terms, as the process is incoherent and is
caused by loss.  Hence it is natural to expect only simple loss terms or
population transfer terms.  The new terms are predicted by this quantum model
because the two exciton modes are coupled together by the phonon interaction,
and so their bosonic nature causes these stimulated effects.  However, because
the phonon scattering is Markovian, the stimulated emission loses its phase
memory and becomes ``stimulated scattering''.  The appearance of these terms
shows clearly the advantages of working from a more rigorous derivation, as
opposed to a phenomenological approach based on simply writing down rate
equations.  They were also predicted by Pau et al \cite{Pau-BJCY-1995} who
started from the Heisenberg equations for this system.

\end{subsection}

\end{section}

\begin{section}{An Ideal Polariton}


A polariton consists of an exciton coupled to a cavity mode.  A full model
would include a description of the different types of exciton, their excited
states, as well as all the cavity modes and the couplings between these. 
However, in practise we know that excitons will only couple to a cavity mode if
they have matching wavevectors.  Also, for low excitation strengths only the
lowest energy exciton is significant.  This means I can consider an ``ideal
polariton'', consisting of an idealised quantum well exciton, an idealised
cavity mode, and the exciton to photon coupling term between the two.  Further,
I use the Coulomb gauge, the dipole approximation, and the rotating wave
approximation to express the coupling term in its simplest form.

  The result couples an exciton to a cavity mode, where both are modelled by
quantised harmonic oscillators.  I write $\bar{e}^\dagger(k)$, $\bar{e} (k)$ as
the operators for the exciton mode, and $\bar{a}^\dagger (k)$, $\bar{a} (k)$ as
the operators for the field in the cavity mode.  The interaction Hamiltonian is

\begin{eqnarray}
\hat{H}_{P} = 
 \hbar A(k) 
    \left[
         \hat{e}^\dagger(k) \hat{a} (k) 
       + \hat{e}(k) \hat{a}^\dagger (k)
    \right] 
\end{eqnarray}

  The parameter $A(k)$ is the coupling between the cavity mode and the exciton.
It is in units of frequency, and is equal to half the splitting between the
pair of polaritons at resonance.  I will denote the exciton energy as $\hbar
\Omega(k)$, and the cavity mode energy as $\hbar \omega(k)$.  This two mode
system of the coupled exciton and cavity mode can be diagonalised, and then be
written in terms of new polariton operators, which describe the two new
polariton modes that can be used in place of the original exciton and photon
modes.  The two eigenvalues (or frequencies) of the eigenmodes (the polaritons)
are
 
\begin{eqnarray}
2 \epsilon_{\pm} (k) 
&=&
   \Omega(k) + \omega(k)
\nonumber \\
&&
 \pm 
   \sqrt{ \left[\Omega(k) - \omega(k) \right] ^2 + 4 A(k) ^2 }
.
\label{p-ideal-energy}
\end{eqnarray}

 The polaritons are characterised by a label $\pm$ denoting the ``upper'' or
``lower'' (in energy) polariton branch, and their wavenumber $k$.  If I introduce
the operators $p_{\pm}^\dagger(k)$, $p_{\pm}(k)$ to describe the polaritons,
the eigenvectors of the system describe how the cavity and exciton operators
combine to form the polariton operators.  The polariton operators for the
coupled exciton - photon system are $\bar{p}_{\pm}^\dagger(k)$,
$\bar{p}_{\pm}(k)$.  The exciton and cavity modes have coupled together to form
two new polariton modes.  

 The polariton operators are related to the exciton and cavity operators 
by the following expressions

\begin{eqnarray}
\hat{p}_{+}(k) 
&=& c_{+}(k) \hat{e} (k) + d_{+}(k) \hat{a} (k)
\nonumber \\
\hat{p}_{-}(k) 
&=& c_{-}(k) \hat{e} (k) + d_{-}(k) \hat{a} (k)
\end{eqnarray}

where

\begin{eqnarray}
 c_{+}(k) 
 &=& - d_{-}(k)
 =  \frac{1}{\sqrt{2}} \sqrt{1 + \sin \left[ \alpha(k) \right] }
\nonumber \\
 d_{+}(k) 
 &=& c_{-}(k)
 =  \frac{-1}{\sqrt{2}} \sqrt{1 - \sin \left[ \alpha(k) \right] }
\nonumber \\
 \tan \left[ \alpha(k) \right] 
 &=& 
   \frac
      { \left[ \Omega(k) - \omega(k) \right] }
      {2 A(k) }
\end{eqnarray}
 
 Note that the equations for the polariton operators in terms of the exciton
and cavity operators can be inverted.  The results are

\begin{eqnarray}
  \hat{e}(k) 
  &=& 
  c_{+}(k) \hat{p}_{+} (k) + c_{-}(k) \hat{p}_{-} (k)
\nonumber \\
  \hat{a}(k) 
  &=& 
  d_{+}(k) \hat{p}_{+} (k) + d_{-}(k) \hat{p}_{-} (k)
.
\label{p-ideal-eadefs}
\end{eqnarray}

 These last equations are the most useful for the work in this paper.  They
allow us to take an interaction involving only the exciton or photon modes and
rewrite it in terms of the eigenstates of the coupled system, ie. in terms of
the polariton operators.

\end{section}
\begin{section}{Coherent driving, losses, and thermal noise}


 Of course it is not only the phonon scattering which is main topic
of this paper that needs to be considered.  Any realistic model 
needs to include losses due to exciton recombination and the imperfect
cavity mirrors.  Also, sources of excitons or polaritons need to be 
included -- these can originate either from the combination of hot 
free carriers to generate a thermal population of excitons, or 
by coherent excitation from a laser beam.
  To do this I use the quantum stochastic differential equation (QSDE) for a
polariton with coherent driving, linear losses and thermal noise.  When
describing a realistic polariton, a sum over in-plane wave number $k$ is needed
to account for the dispersion. However, since each $k$ mode is independent in
these interactions, I do not need to include the sum over $k$ states here.  The
standard driving and loss terms for a boson mode described by operators
$\hat{f}$, $\hat{f} ^\dagger$ have an interaction Hamiltonian like

\begin{eqnarray}
\hat{H}_D
=
\imath
\left[
   \epsilon ^* \hat{f} - \epsilon \hat{f} ^\dagger 
\right]
+
\sum_i
\gamma_i '
   \left[
      \hat{f} \hat{\Gamma}_i ^\dagger
      +
      \hat{f} ^\dagger \hat{\Gamma}_i 
   \right]
.
\end{eqnarray}

 The first term describes a coherent driving process, which can be imagined as
being due to a classical driving field, but is actually a coherent state 
driving term. 
The second term couples the mode to a reservoir consisting of an infinite
number of boson modes $\hat{\Gamma}_i ^\dagger$, $\hat{\Gamma}_i$. 
The loss from the system is modelled by irreversible transfer of 
excitation from the system into this reservoir.
The amount of loss is described by a decay rate $\gamma$ that is
related to the $\gamma_i '$, but includes the sum over the density of states of
the reservoir.

 In a polariton system the excitons are not the eigenstates. The eigenstates
are the polaritons, and so I need to replace the exciton operators with their
equivalent in terms of polariton operators, defined by eqn
(\ref{p-ideal-eadefs}).  I will now use $f_{\pm}$ to represent $c_{\pm}$ or
$d_{\pm}$, depending on whether I am considering interaction with the cavity
mode ($\hat{f} = \hat{a}$) or the exciton ($\hat{f} = \hat{e}$).  In polariton
operators, the interaction Hamiltonian can be rewritten 

\begin{eqnarray}
\hat{H}_D
&=&
\imath
\left\{
   \left[ \epsilon ^* f_{+} \right]
   \hat{p}_{+} (t) 
   -
   \left[ \epsilon ^* f_{+} \right] ^*
   \hat{p}_{+} ^\dagger (t) 
\right\}
\nonumber \\
&&
 +
\imath
\left\{
   \left[ \epsilon ^* f_{-} \right]
   \hat{p}_{-} (t) 
   -
   \left[ \epsilon ^* f_{-} \right] ^*
   \hat{p}_{-} ^\dagger (t) 
\right\}
\nonumber \\
& &
+
\sum_i
   \left[
      f_{+}     \hat{p}_{+}          (t) \hat{\Gamma}_i ^\dagger 
    + f_{+} ^*  \hat{p}_{+} ^\dagger (t) \hat{\Gamma}_i 
   \right]
\nonumber \\
&&
+
\sum_i
   \left[
      f_{-}     \hat{p}_{-}          (t) \hat{\Gamma}_i ^\dagger 
    + f_{-} ^*  \hat{p}_{-} ^\dagger (t) \hat{\Gamma}_i 
   \right]
.
\end{eqnarray}

The QSDE for an arbitrary system operator $\hat{A}$ affected by this
Hamiltonian is just \cite{Gardiner-qn}.

\begin{eqnarray}
&&\frac{d}{dt} \hat{A}
=
\left[ 
   \left\{ \epsilon ^* f_{+} \right\}
   \hat{p}_{+} 
   -
   \left\{ \epsilon ^* f_{+} \right\} ^*
   \hat{p}_{+} ^\dagger 
 ,
  \hat{A}
\right] 
dt
\nonumber \\
&&
+
\frac{\gamma}{2}
\left[ N + 1 \right] \left[ f_{+} ^* f_{+} \right]
\left[
   2 \hat{p}_{+} ^\dagger \hat{A} \hat{p}_{+} 
- 
   \hat{A} \hat{p}_{+} ^\dagger \hat{p}_{+} 
- 
   \hat{p}_{+} ^\dagger \hat{p}_{+} \hat{A} 
\right]
\nonumber \\
& &
+
\frac{\gamma}{2}
\left[ N \right] 
\left[ f_{+} ^* f_{+} \right] 
\left[
   2 \hat{p}_{+} \hat{A}  \hat{p}_{+} ^\dagger 
- 
   \hat{A} \hat{p}_{+} \hat{p}_{+} ^\dagger 
- 
   \hat{p}_{+} \hat{p}_{+} ^\dagger \hat{A} 
\right]
\nonumber \\
&&
+ 
   \sqrt{ \gamma } 
   \left[ \hat{A}, f_{+} ^* \hat{p}_{+} ^\dagger \right] 
   d \hat{B}
+ 
   \sqrt{ \gamma } 
   d \hat{B} ^\dagger
   \left[ \hat{A}, f_{+} \hat{p}_{+} \right] 
\nonumber \\
&&
+ \left[ ... \right]
.
\end{eqnarray}

 The $\left[ ... \right]$ represent the $\hat{p}_{-}$ terms, which are similar
to the $\hat{p}_{+}$ ones, and $N$ is the thermal occupation number of the
reservoir modes.  Different sets of reservoir modes will have different thermal
populations according to a Boltzmann weight factor.  A reservoir coupled to a
mode of frequency $\omega$ will have a thermal occupation $N$ proportional to
$\exp\left[ - \hbar \omega / k_B T \right]$.  At optical frequencies and room
temperature, this factor is so small that there are usually no significant
thermal effects.  The quantum noise operators $d\hat{B}$, $d\hat{B}^\dagger$
are defined in the same way as those in eqn (\ref{e-phonon-qnoise}).

\begin{subsection}{Intensity and amplitude equations}

To get a clear picture of what these interactions do, I will
calculate some specific examples of these QSDE's.  In the quasi-PL and coherent
excitation models presented in section VI, equations for excitation number and
mode amplitude are used; more specifically, these are derived from the QSDE's
for the mode intensity and mode amplitude operators.  First I consider the
equation of motion for polaritons with the exciton (or photon) affected by
losses and thermal noise, but with no coherent driving term.  The QSDE for how
the intensity $\hat{I}_{\pm} = \hat{p}_{\pm} ^\dagger \hat{p}_{\pm} $ of the
mode changes due to the losses and the associated thermal noise  is 
 
\begin{eqnarray}
\frac{d}{dt} \hat{I}_{\pm}
&=&
\gamma f_{\pm} ^* f_{\pm}
\left[
   - \hat{I}_{\pm} +  N 
\right]
dt
\nonumber \\
&&
-
\sqrt{\gamma}
\left[
f_{\pm} ^* \hat{p}_{\pm} ^\dagger d\hat{B}
+
f_{\pm} d\hat{B} ^\dagger \hat{p}_{\pm}
\right]
.
\end{eqnarray}

 Rate equations can be produced from this QSDE easily.  All we need to do is to
neglect the quantum noise terms, and replace the intensity operators with
numbers.  This is done more rigorously by taking the expectation values of the
equation, with the quantum noise terms averaged to zero.  Replacing the
expectation value with a number $I_{\pm} = \left< \hat{I}_{\pm} \right>$, the
equation can be rewritten as 

\begin{eqnarray}
\frac{d}{dt} I_{\pm}
&=&
\gamma f_{\pm} ^* f_{\pm}
\left[
   - I_{\pm} + N
\right]
.
\end{eqnarray}

  Now I consider the effect of a coherent driving term as well as losses.  The
QSDE for how the operator $\hat{p}_{\pm}$ of the mode changes due to the losses
and the associated noise is 

\begin{eqnarray}
d \hat{p}_{\pm}
&=&
\left[
   f_{\pm} ^* \epsilon 
- 
   f_{\pm} ^* f_{\pm} \frac{\gamma}{2} \hat{p}_{\pm}
\right]
dt
+
\sqrt{\gamma} f_{\pm} ^* d\hat{B}
.
\end{eqnarray}

  A complex number equation for the mode amplitude can be produced from this
QSDE easily.  We can take the expectation values of the equation, and assume
the mode is in a coherent state $\left| \alpha \right>$.  The quantum noise
term then averages to zero, and we get an equation in terms of the coherent 
amplitude $\alpha$ 

  While this approach might seem fine if the mode is in a coherent state, what
if it is not?  Fortunately, the basic idea can still be used.  Any quantum
state can be expanded over a basis of coherent states \cite{Gardiner-qn}, by
using either the Glauber-Sudarshan P representation
\cite{Glauber-1963,Sudarshan-1963}, or by the subsequent generalisations
\cite{Drummond-G-1980}.  Here the systematic procedure is to convert the QSDE
into the equivalent master equation form and then apply the representations in
the standard way.  In practise this means that the state of the mode is not
specified by a single complex number $\alpha$, but by an appropriately
distributed ensemble, each member of which follows the equation

\begin{eqnarray}
\frac{d}{dt} \alpha
&=&
   f_{\pm} ^* \epsilon 
- 
   f_{\pm} ^* f_{\pm} \frac{\gamma}{2} \alpha
.
\end{eqnarray}

  In these coherent state representations the non thermal part of the quantum
noise terms vanish, as they are absorbed into the coherent state basis.  The
effect of thermal noise on the system is to spread the ensemble out -- this
does not affect its average amplitude, but will change its intensity by some
amount.  If thermal noise can be neglected, and the system starts or is 
being coherently driven, then it can be adequately described by a single 
complex number.  However, if there are additional interactions, such as 
strong phonon scattering, then the system might well depart strongly from 
a coherent state.

\end{subsection}

\end{section}

\begin{section}{The polariton-phonon interaction}


 In section II I described an interaction Hamiltonian for excitons scattering
off phonons. To convert it into one relevant to a polariton system, I employ
the same procedure as for the coherent driving and decay interactions.  Since
the eigenstates of the system are the polaritons, and so I need to replace the
exciton operators with their equivalent in terms of polariton operators, as
defined by eqn (\ref{p-ideal-eadefs}).  A single exciton and cavity mode branch
gives us a two polariton system, so the interaction Hamiltonian in the
polariton operators is

\begin{eqnarray}
&\hat{H}&
=
\sum_{k} \sum_{\Delta k} \sum_q
   \chi(k, k-\Delta k; q)
\nonumber \\
& &
\left\{
   \left[
      d_{+} (k) \hat{p}_{+} (k; t) 
    + d_{-} (k) \hat{p}_{-} (k; t)
   \right]
\right.
\nonumber \\
&&
\left.
   \left[
      d_{+} ^* (k - \Delta k) \hat{p}_{+} ^\dagger (k - \Delta k; t) 
   \right.
\right.
\nonumber \\
&&
\left.
   \left.
    + d_{-} ^* (k - \Delta k) \hat{p}_{-} ^\dagger (k - \Delta k; t)
   \right]
   \hat{b} ^\dagger (\Delta k; q)
\right.
\nonumber \\
& &
\left.
  +
   \left[
      d_{+} ^* (k) \hat{p}_{+} ^\dagger (k; t) 
    + d_{-} ^* (k) \hat{p}_{-} ^\dagger (k; t)
   \right]
\right.
\nonumber \\
&&
\left.
   \left[
      d_{+} (k - \Delta k) \hat{p}_{+} (k - \Delta k; t) 
   \right.
\right.
\nonumber \\
&&
\left.
   \left.
    + d_{-} (k - \Delta k) \hat{p}_{-} (k - \Delta k; t)
   \right]
   \hat{b} (\Delta k; q)
\right\}
.
\end{eqnarray}

 When this is expanded out, there are four types of important term. Two are
intra- polariton terms, involving phonon scattering along the upper or lower
polariton branches, and the other two are inter- polariton scatterings.  In
addition to these, there are non-resonant terms which do not conserve energy
such as those that create or destroy two polaritons.

  How do all these processes fit together?  First consider the dispersion
curves for the polaritons and the acoustic phonons.  This is a 2D problem --
the polariton is confined in the growth direction by the microcavity DBR
mirrors, and the quantum well.  Because both polaritons only have in-plane
freedom, only phonons that conserve in-plane momentum can be emitted.  The
dispersion relations for the polaritons are bowl-like surfaces, and the phonon
dispersion is like a solid cone.  Acoustic phonons carry very little energy for
their momentum, so this cone is very shallow compared to the other dispersions.
The 'cone' is solid because the phonons can have a non zero growth direction
momentum -- the internal parts of the cone correspond to when some of the 
phonon momentum is in the growth direction.

\begin{figure}
\includegraphics[width=80mm]{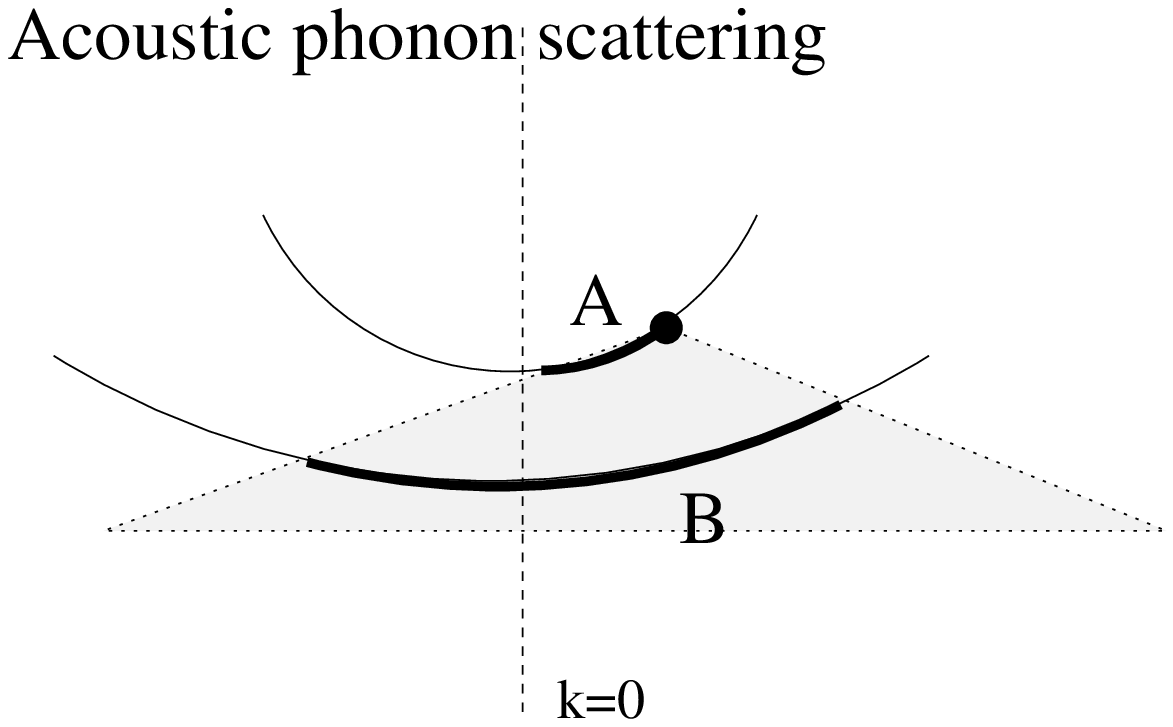}
\caption{ 
\label{Graph-AP}
{ 
The acoustic phonon scattering process.  The thin curves
represent the dispersion relations defining the exciton or polariton branches. 
The triangular shaded area gives the allowed energies and in-plane momenta of
the phonons.  The area is shaded because the phonons can have a growth
direction momenta which has been projected out.  The overlap of the branches 
and the shaded area is
the set of possible final exciton or polariton states given that the initial
state was at the dot at the apex of the phonon triangle. (A) Intra-polariton 
phonon
scattering. The emission of a phonon can transfer the polariton from its
initial state to another part of the polariton branch.  (B) Inter-polariton
phonon scattering.  The emission of a phonon can transfer the polariton from
its initial state to part of the lower polariton branch. 
}
}
\end{figure}

\begin{figure}
\includegraphics[width=80mm]{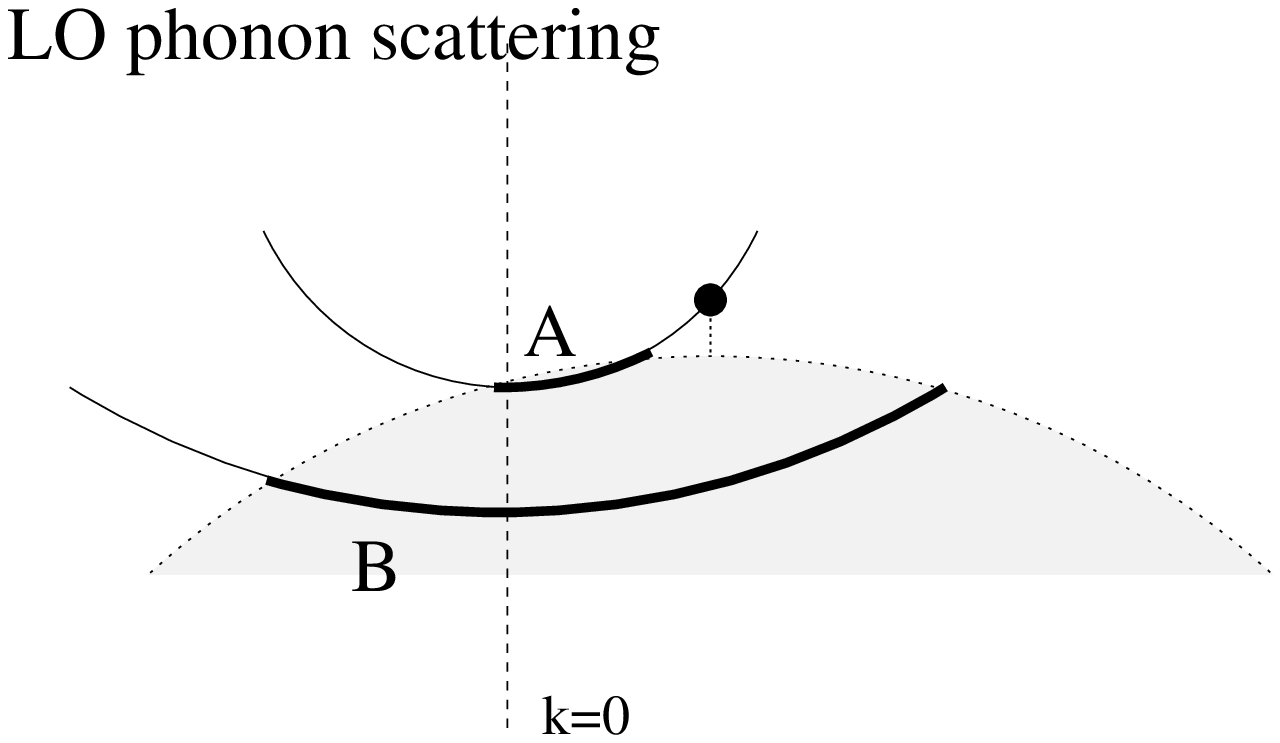}
\caption{ 
\label{Graph-OP}
{ 
The LO phonon scattering process.  The thin curves
represent the dispersion relations defining the exciton or polariton branches. 
The roughly parabolic shaded area gives the allowed energies and in-plane
momenta of the LO phonons.  The area is shaded because the phonons can have a
growth direction momenta which has been projected out.  The top of the parabola
is displaced downwards from the dot denoting the initial position of the 
exciton or polariton by the minimum energy of an LO phonon.  The overlap of
the branches and the shaded area is the set of possible final exciton or 
polariton states given that
the initial state was above the apex of the phonon parabola. (A) and (B) show
the intra and inter-polariton scattering.  
}
}
\end{figure}

  Figure \ref{Graph-AP} is a schematic diagram showing the dispersion curves
and allowed acoustic phonon scattering processes.  The intra-branch process
mainly transfers population from large $k$ to small $k$ along the same branch,
although the opposite can also occur in the presence of thermal phonons.  The
inter-branch processes transfers population between the two branches,
predominantly from the upper $({+})$ to the lower $({-})$ branch.  Again, the
reverse process is possible in the presence of thermal phonons.  However, note
that the upwards thermal processes use a phonon dispersion with its apex on the
lower initial state.  Similarly, figure \ref{Graph-OP} shows the
dispersion curves and allowed optical phonon scattering processes.  This
follows the same principle as the acoustic case, but with the optical phonon
dispersion curve and so includes a minimum phonon energy.

 Here I am primarily concerned with the downward inter-branch process, from the
upper to the lower branch with the production of a phonon.  This means a RWA
can be justified for sufficiently large branch separation 
as the non resonant terms will oscillate at a rate given by twice 
this -- much faster than the speed of 
the important system dynamics. The resonant upper to lower phonon branch  
interaction term is then contained in 

\begin{eqnarray}
&\hat{H}&
=
\sum_{k} \sum_{\Delta k} \sum_q
   \chi(k, k-\Delta k; q)
\nonumber \\
& &
   \left[
         d_{+} (k) \hat{p}_{+} (k; t) 
         d_{-} ^* (k - \Delta k) \hat{p}_{-} ^\dagger (k - \Delta k; t)
         \hat{b} ^\dagger (\Delta k; q)
   \right.
\nonumber \\
&&
   \left.
     +
         d_{+} ^* (k) \hat{p}_{+} ^\dagger (k; t)
         d_{-} (k - \Delta k) \hat{p}_{-} (k - \Delta k; t)
      \hat{b} (\Delta k; q)
   \right]
\nonumber \\
.
\end{eqnarray}

 Note that this is the same as the exciton phonon Hamiltonian if we replace 
$d_{+} (k) \hat{p}_{+} (k)$ with $\hat{e} (k)$ and $d_{-} ^* (k - \Delta k)
\hat{p}_{-} ^\dagger (k - \Delta k)$ with $\hat{e} (k - \Delta k)$.

\begin{subsection}{The polariton-phonon QSDE}

 The justification for the validity of the QSDE method for this system is
identical to that used in the initial exciton scattering section, so I do not
repeat it here.  The only thing that has changed is the introduction of the
coefficients for the component of exciton in the polariton, and the particular
allowed scattering transitions between branches.  The Ito QSDE for a general
system operator $\hat{S}$ in a two polariton system undergoing $({+})$ to
$({-})$ phonon scattering is

\begin{eqnarray}
&&d \hat{S}
=
\sum_{\Delta k}
   \frac{ \chi(k, k-\Delta k) }{2} 
   \left| d_{+} (k) d_{-} ^* (k - \Delta k) \right| ^2
\nonumber \\
& & ~~~~~
\left\{
   \left( N_P (\Delta k) + 1 \right)
\right.
\nonumber \\
&&
\left.
   \left[
      2 
      \hat{p}_{+} ^\dagger (k) \hat{p}_{-} (k - \Delta k)
      \hat{S}
      \hat{p}_{+} (k) \hat{p}_{-} ^\dagger (k - \Delta k)
   \right.
\right.
\nonumber \\
& &
\left.
   \left.
     -
      \hat{p}_{+} ^\dagger (k) \hat{p}_{-} (k - \Delta k) 
      \hat{p}_{+} (k) \hat{p}_{-} ^\dagger (k - \Delta k) 
      \hat{S}
   \right.
\right.
\nonumber \\
& &
\left.
   \left.
     -
      \hat{S}
      \hat{p}_{+} ^\dagger (k) \hat{p}_{-} (k - \Delta k) 
      \hat{p}_{+} (k) \hat{p}_{-} ^\dagger (k - \Delta k) 
   \right] dt
\right.
\nonumber \\
&&
\left.
 +
   N_P (\Delta k)
   \left[
      2 
      \hat{p}_{+} (k) \hat{p}_{-} ^\dagger (k - \Delta k)
      \hat{S}
      \hat{p}_{+} ^\dagger (k) \hat{p}_{-} (k - \Delta k)
   \right.
\right.
\nonumber \\
& &
\left.
   \left.
     -
      \hat{p}_{+} (k) \hat{p}_{-} ^\dagger (k - \Delta k) 
      \hat{p}_{+} ^\dagger (k) \hat{p}_{-} (k - \Delta k) 
      \hat{S}
   \right.
\right.
\nonumber \\
&&
\left.
   \left.
     -
      \hat{S}
      \hat{p}_{+} (k) \hat{p}_{-} ^\dagger (k - \Delta k) 
      \hat{p}_{+} ^\dagger (k) \hat{p}_{-} (k - \Delta k) 
   \right] dt
\right\}
\nonumber \\
& &
-
\sum_{\Delta k}
\sqrt{ \chi(k, k-\Delta k) }
   \left[ d_{+} ^* (k) d_{-} (k - \Delta k) \right]
\nonumber \\
&&
   \left[ \hat{S}, \hat{p}_{+} ^\dagger (k) 
          \hat{p}_{-} (k - \Delta k) \right]
   d \hat{B} 
\nonumber \\
& &
+
\sum_{\Delta k}
\sqrt{ \chi(k, k-\Delta k) }
\left[ d_{+} (k) d_{-} (k - \Delta k) ^* \right]
\nonumber \\
&&
   d \hat{B} ^\dagger 
   \left[ \hat{S}, \hat{p}_{+} (k) 
          \hat{p}_{-} ^\dagger (k - \Delta k) \right]
.
\end{eqnarray}

  The operators $d \hat{B}_P$, $d \hat{B}_P ^\dagger$ are quantum noise
operators which replace the infinity of phonon operators $\hat{b}$ and 
$\hat{b} ^\dagger$,
and $N_P(\Delta k)$ is the average number of thermal phonons in each
phonon mode.  The summation over $q$ has dropped out because the resonance
condition allows us to calculate it in terms of the $k$ terms and the
dispersion curves.  The noise operators $d\hat{B}_P$, $d\hat{B}_P ^\dagger$ are
defined in the same way as in eqn (\ref{e-phonon-qnoise}).

\end{subsection}

\begin{subsection}{Intensity and amplitude operator equations}

  Similarly to subsection II.B, I calculate some relevant examples of these
QSDE's, and convert them into rate and complex number equations.  For
simplicity, I write $\hat{p}_{+} (k)$ as $\hat{p}_{+}$, and $\hat{p}_{-}
(k-\Delta k)$ as $\hat{p}_{-}$. The intensity for the $({+})$ mode is just
$\hat{I}_{+} = \hat{p}_{+} ^\dagger \hat{p}_{+}$, that for the $({-})$ mode is
$\hat{I}_{-} = \hat{p}_{-} ^\dagger \hat{p}_{-}$. The polariton coefficients I
will also simplify, writing $d_{+} (k)$ as $d_{+}$, and $d_{-} (k- \Delta k)$
as $d_{-}$, and $\chi (k, k- \Delta k)$ as simply $\chi$.

  The QSDE equations for the intensity of the $({+})$ and $({-})$ modes due to
the phonon scattering are

\begin{eqnarray}
d \hat{I}_{+}
&=&
\chi 
\left| d_{+} d_{-} ^* \right| ^2
\left[ 
- \hat{I}_{+} \left( \hat{I}_{-} + 1 \right) 
+  N_P
   \left( \hat{I}_{-} - \hat{I}_{+} \right)
\right] dt
\nonumber \\
&&
- \sqrt{ \chi }
\left[
   d_{+} d_{-} ^* \hat{p}_{+} ^\dagger \hat{p}_{-} d\hat{B}_P 
 + d_{+} ^* d_{-} d\hat{B}_P ^\dagger \hat{p}_{+} \hat{p}_{-} ^\dagger
\right]
,
\nonumber \\
d \hat{I}_{-}
&=&
\chi 
\left| d_{+} d_{-} ^* \right| ^2
\left[ 
   \hat{I}_{+} \left( \hat{I}_{-} + 1 \right) 
-  N_P
   \left( \hat{I}_{-} - \hat{I}_{+} \right)
\right] dt 
\nonumber \\
&&
+ \sqrt{\chi} 
\left[
   d_{+} ^* d_{-} \hat{p}_{+} ^\dagger \hat{p}_{-} d\hat{B}_P 
 + d_{+} d_{-} ^* d\hat{B}_P ^\dagger \hat{p}_{+} \hat{p}_{-} ^\dagger
\right]
.
\end{eqnarray}

 As in the exciton case, the noise terms cannot be rewritten in a form
depending solely on $\hat{I}_{\pm}$, so I leave them in terms of the
$\hat{p}_{+}$, $\hat{p}_{-}$.  Again the terms in the pair of equations balance
each other to preserve excitation number.  Using the same procedure as in IV.A,
these have a rate equation form

\begin{eqnarray}
\frac{d}{dt} I_{+}
&=&
\chi 
\left| d_{+} d_{-} ^* \right| ^2
\left[ 
- I_{+} \left( I_{-} + 1 \right) 
+  N_P
   \left( I_{-} - I_{+} \right)
\right]
,
\nonumber \\
\frac{d}{dt}  I_{-}
&=&
\chi 
\left| d_{+} d_{-} ^* \right| ^2
\left[ 
   I_{+} \left( I_{-} + 1 \right) 
-  N_P
   \left( I_{-} - I_{+} \right)
\right]
.
\end{eqnarray}

Similarly, the QSDE equations for the amplitude of the $({+})$ and $({-})$
modes due to the phonon scattering are

\begin{eqnarray}
d \hat{p}_{+}
&=&
\chi 
\left| d_{+} d_{-} ^* \right| ^2
- \hat{p}_{+} 
   \left[ 
      \hat{p}_{-} ^\dagger \hat{p}_{-} + 1 + N_P   
   \right] dt
\nonumber \\
&&
+ \sqrt{\chi} d_{+} ^* d_{-} \hat{p}_{-} d \hat{B}_P
,
\nonumber \\
d \hat{p}_{-}
&=&
 \chi 
 \left| d_{+} d_{-} ^* \right| ^2
 \hat{p}_{-}
 \left[ 
    \hat{p}_{+} ^\dagger \hat{p}_{+} - N_P 
 \right] dt 
\nonumber \\
&&
+ 
 \sqrt{\chi} d_{+} d_{-} ^* d\hat{B}_P ^\dagger \hat{p}_{+} 
.
\end{eqnarray}

 Using the procedure referred to in section IV.A, these have a complex 
number form for the coherent amplitude, which is

\begin{eqnarray}
d \alpha_{+}
&=&
\chi 
\left| d_{+} d_{-} ^* \right| ^2
\left[ 
- \alpha_{+} \left( \alpha_{-} ^\dagger \alpha_{-} + 1 \right) 
- N_P \alpha_{+} 
\right] dt
\nonumber \\
&&
+ \sqrt{ 2 \chi d_{+} ^* d_{-} N_P \alpha_{-} ^\dagger \alpha_{-} } dW_A
\nonumber \\
&&
+ \imath 
  \sqrt{ \chi d_{+} ^* d_{-} ( 1 + 2 N_P ) \alpha_{+} \alpha_{-}  } dW_2
,
\nonumber \\
d \alpha_{-}
&=&
\chi 
\left| d_{+} d_{-} ^* \right| ^2
\left[ 
   \alpha_{+} ^\dagger \alpha_{+} \alpha_{-}
-  N_P \alpha_{-}
\right] dt 
\nonumber \\
&&
+ \sqrt{ 2 \chi d_{+} ^* d_{-} ( 1 + N_P ) \alpha_{+} ^\dagger \alpha_{+} } dW_B
\nonumber \\
&&
+ \imath 
  \sqrt{ \chi d_{+} ^* d_{-} ( 1 + 2 N_P ) \alpha_{+} \alpha_{-} } dW_2 ^*
.
\end{eqnarray}

These polariton phonon scattering equations have the same features
as the exciton equations derived in section II.B.  This means the same 
comments apply here too -- notably those about stimulated scattering.
The new terms $dW$ are white noise increments, $dW_A$ and $dW_B$ are 
uncorrelated real white noises, and $dW_2$ is a complex white noise, 
where ($i=\left\{ A, B, 2 \right\}$)

\begin{eqnarray}
\left< dW_i(t) dW_j(t')^* \right>
=
\delta_{ij} \delta(t-t')
.
\end{eqnarray}

\end{subsection}

\end{section}


\begin{section}{Polariton phonon models}


  Here I consider two models -- one of a photoluminescence-like process and the
other involving coherent excitation.  Both models involve a pair of polariton
branches that result from the coupling of a single exciton and a cavity.  The
polaritons are described by boson operators $\hat{p}_{\pm} (k)$, $\hat{p}_{\pm}
^\dagger (k)$, indexed by in-plane wave vector.  The splitting between the
higher $({+})$ and lower $({-})$ energy polaritons is assumed to be less than
the minimum optical phonon energy, so only acoustic phonons need to be
included.  The coupling to acoustic phonons is much weaker than that to optical
phonons, and this causes the bottleneck effect.

  The photon component of the polaritons can decay due to cavity losses, which
I model by coupling to optical reservoirs.  These represent the field modes 
outside the cavity.  The cavity decay rate is $\gamma_L
(k)$, with thermal photon number $N_L (k)$, and quantum noise operators $d
\hat{B}_L (k)$, $d \hat{B}_L ^\dagger (k)$.  The exciton component of 
polaritons on either branch can recombine and emit photons, which I 
model by coupling each exciton mode to independent photon 
reservoirs.  This recombination rate is $\gamma_E (k)$, with thermal
photon number $N_E (k)$, and quantum noise operators $d \hat{B}_E (k)$, $d
\hat{B}_E ^\dagger (k)$.

  The acoustic phonon process couples the $({+})$ modes to all allowed $({-})$
modes.  The presence of thermal phonons is described using a thermal phonon
number $N_P (k)$.  Experiments done by Fainstein et al \cite{Fainstein-JT-1995}
for Raman scattering but in an otherwise similar situation show no power
dependence in the spectra.  This indicates that the effects of thermal phonons
are likely to be negligible, so we can set $N_P (k) = 0$.  Only the downwards
$({+}) \rightarrow ({-})$ inter-polariton phonon scattering process is included
here.
 
 I will simplify both models with some physical considerations. 
Both the recombination reservoirs and the cavity decay reservoirs have very low
thermal excitations ($N_L (k), N_E (k) \approx 0$), because the exciton and
cavity mode energies are much greater than the thermal energy.  Further, the
recombination decay $\gamma_E (k)$ is typically a much slower process than the
cavity decay rate $\gamma_E (k) \ll \gamma_L (k)$, and so can usually be 
ignored when the
polariton has appreciable fractions of both cavity mode and exciton.

  In the following equations, I will suppress the $k$ argument when $k=0$ for
brevity, so that $\hat{p}_{+} = \hat{p}_{+} (0)$, $c_{\pm} = c_{\pm} (0)$,
$d_{\pm} = d_{\pm} (0)$, $\gamma_L = \gamma_L (0)$, $\gamma_E = \gamma_E (0)$,
$N_L = N_L (0)$, $N_E = N_E (0)$, and I also write $\chi (0, k)$ as $\chi (k)$.

\begin{subsection}{Photoluminescence}


 The photoluminescence model is one in which the polariton is pumped as the
free carriers excited by the driving field can combine in concert with the
cavity mode to form polaritons.  Exciton formation in PL is a complicated
process, with free carriers combining to form hot excitons (with $k \gg 0$),
and excited excitons (eg. those with extra orbital momentum).  These decay by
LO or acoustic phonon emission or other scattering processes into the ``cold''
exciton states that can decay radiatively, and can therefore be seen optically.
The situation with polaritons is similar.

  A full description of this process is beyond the scope of this paper, so here
I use an approximate model.  I suppose that the recombination processes
preferentially fill the upper polariton branch, something which could occur
because the upper branch has higher energy than the lower branch for the 
same $k$.  This agrees wth the results that originally motivated this paper 
\cite{Tribe-etal} in which the upper branch is preferentially populated.    
Further, I assume the statistical properties of the newly created polaritons
are influenced most strongly by the thermal nature of the free carriers they
originated from.  We suppose that the predominant effect of the optical pumping
is that each mode ($k$ value) of the upper polariton branch is driven by an
independent thermal population, modelled by coupling to a set of reservoir
modes which cause a weak loss $\gamma_C (k) \approx 0$, but have a high thermal
occupancy $N_C (k)$ so there is significant thermal driving strength $M_C (k) =
\gamma_C (k) N_C (k)$.  The free carrier decay rate is small because the
polaritons are unlikely to be scattered into free carriers.  

  I make the further simplifying assumption that the finite $k$ higher 
polariton states decay rapidly down into the $k=0$ state due to 
optical and acoustic phonon
scattering.  As a result only the $k=0$ upper polariton state is relevant, 
and I couple a single $k=0$ upper polariton to a
reservoir described by the parameters $\gamma_C$, $M_C$, and the quantum noise
operators $d\hat{B}_P$, $d\hat{B}_P ^\dagger$.  

The complete equations for $\hat{I}_{+} = \hat{I}_{+} (0) $, $\hat{I}_{-} (k)$
can then be constructed.  Firstly, however, I can simplify them with the
physical considerations given above.  For the $({+})$ mode the sources of
thermal noise are swamped by the free carrier component, since $M_C = \gamma_C
N_C \gg \gamma_L (k) N_L (k), \gamma_E (k) N_E (k)$.  Also, the free carrier
decay rate is much smaller than the cavity decay rate 
$\gamma_C \ll \gamma_L (k)$.  These simplifications hold true for 
all cases where the polariton has
appreciable fractions of both cavity mode and exciton.  

  The phonon scattering itself is very weak, so that $ \gamma_L \gg \sum_k \chi
(k) \left| d_{+} d_{-} ^* (k) \right| ^2 $.  In the typical situation where
$\hat{I}_{-} (k)$ remains small because the lower polariton is only populated
by this scattering process, this means that the effect of the phonon scattering
terms in the equation for $\hat{I}_{+}$ is negligible.

\begin{figure}
\includegraphics[width=80mm]{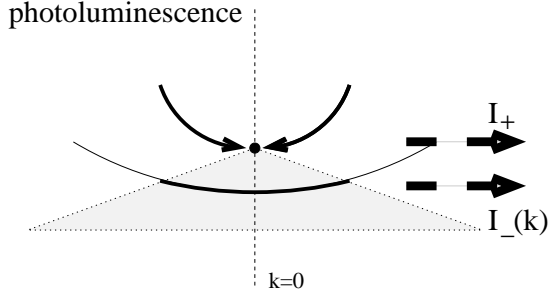}
\caption{ 
\label{Graph-PL}
{
The quasi- photoluminesence model.  The thin curves are
the polariton branches.  The triangular shaded area gives the allowed 
energies and in-plane momenta of the acoustic phonons.  The overlap of 
these two is the set of
possible final states given that the initial state was at $k=0$. The thick
arrows represent excitation cascading down the upper branch by primarily LO
phonon scattering to give a thermal population near the $k=0$ upper state.  The
thick lower curve is the allowed final states after emission of an acoustic
phonon.  The dashed arrows are the light emitted by decay from the polaritons
through the cavity mode.
}
}
\end{figure}

 Figure \ref{Graph-PL} shows a schematic diagram of the model.  The QSDE's for
it are

\begin{eqnarray}
d \hat{I}_{+}
&\approx&
\left\{
- \gamma_L \left| c_{+} \right| ^2 \hat{I}_{+}
+  M_C 
\right\} dt
\nonumber \\
&&
+ 
   \sqrt{ \gamma_L } c_{+} ^* \hat{p}_{+} ^\dagger d \hat{B}_L (0) 
\nonumber \\
&&
 + \sqrt{ \gamma_L } c_{+}    d \hat{B}_L (0) ^\dagger \hat{p}_{+} 
,
\\
d \hat{I}_{-} (k)
&\approx&
\left\{
-     \gamma_L (k) \left| c_{-} (k) \right| ^2 \hat{I}_{+}
\right.
\nonumber \\
&&
\left.
+
   \chi (k) \left| d_{+} d_{-} ^* (k) \right| ^2
   \hat{I}_{+} \left[ \hat{I}_{-} (k) + 1 \right] 
\right\} dt
\nonumber \\
&&
+ 
   \sqrt{ \gamma_L (k) } c_{-} ^* (k) 
   \hat{p}_{-} ^\dagger (k) d \hat{B}_L (k) 
\nonumber \\
& &
+ 
   \sqrt{ \gamma_L (k) } c_{-}    (k) 
   d \hat{B}_L (k) ^\dagger     \hat{p}_{+} (k)
\nonumber \\
&&
+
   \sqrt{ \chi(k) } d_{+} ^* d_{-} (k) 
   \hat{p}_{+} ^\dagger \hat{p}_{-} (k) 
   d \hat{B}_P (0)  
\nonumber \\
&&
+ 
   \sqrt{ \chi(k) } d_{+} d_{-} ^* (k) 
   d \hat{B}_P ^\dagger (0)
   \hat{p}_{+} \hat{p}_{-} ^\dagger (k)
.
\end{eqnarray}

  Since in this application we do not need the detailed quantum mechanical
noise correlations, we can approximate them by taking the expectation values
and factorising the different intensity moments (see section IV.A).
Replacing the expectation value with a number 
$I_{\pm} = \left< \hat{I}_{\pm} \right>$, 
I can describe the system using rate equations --

\begin{eqnarray}
\frac{d}{dt} I_{+}
&\approx&
\left[ 
-  \gamma_L (0)
   \left| c_{+} \right| ^2
    I_{+}
+
   M_C
\right]
,
\nonumber \\
\frac{d}{dt} I_{-} (k)
&\approx&
\left[ 
-  \gamma_L (k) 
   \left| c_{-} (k) \right| ^2
    I_{-} (k)
\right.
\nonumber \\
&&
\left.
+  
   \chi (k) 
   \left| d_{+} d_{-} (k) \right| ^2
   I_{+}
   \left[ I_{-} (k) +1 \right]
\right]
.
\end{eqnarray}

The exciton modes have steady state populations of 

\begin{eqnarray}
I_{+} 
&=&
\frac
   { M_C }
   { \left| c_{+} \right| ^2 \gamma_L }
,
\nonumber \\
I_{-} (k) 
&\approx& 
\frac
{ \chi (k) \left| d_{+} d_{-} (k) \right| ^2 I_{+}  }
{ \left| c_{-} (k) \right| ^2 \gamma_L (k) }
,
\end{eqnarray}

where the equation for $I_{-} (k)$ has been further simplified by reusing the
fact that the phonon scattering is a weak process, with $\left| c_{-} (k)
\right| ^2 \gamma_L (k) \gg \sum_k \chi (k) \left| d_{+} d_{-} (k) \right| ^2$
and $I_{+}$ small.  Each polariton mode emits photons at a rate given by the
product of its decay rate $ \left| c_{\pm} (k) \right| ^2\gamma_L (k)$ and the
occupation $I_{\pm}$.  For the upper $({+})$ mode this is just $M_C$.  Using
the $({+})$ mode population, the $({-})$ modes have populations of

\begin{eqnarray}
I_{-} (k) 
&=& 
\frac
{ \chi (k) \left| d_{+} d_{-} (k) \right| ^2 M_C }
{ \left| c_{+} \right| ^2 \gamma_L 
  \left| c_{-} (k) \right| ^2 \gamma_L (k) 
}
.
\end{eqnarray}

The ratio of cavity decay intensities between all $({-})$ modes and the $k=0$
$({+})$ mode is 

\begin{eqnarray}
R
&=& 
\sum_k
\frac
 { \chi (k) \left| d_{+} d_{-} (k) \right| ^2 }
 { \left| c_{+} \right| ^2 \gamma_L }
.
\end{eqnarray}

From this expression we can see that the relative strengths of the upper and
lower polariton emission peaks is controlled largely by the strength of the
phonon scattering between the two branches. As a result, the upper peak is
larger than the lower peak. In contrast, a model that coupled the free carriers
to both branches would populate them according to their Boltzmann factors,
and would have the lower exciton peak as the larger because of its greater
thermal population.

\end{subsection}

\begin{subsection}{Coherent excitation}


  Now I will now consider a system where the higher energy polariton $({+})$ is
coherently excited by a laser beam tuned to $k=0$.  The laser field is assumed
to be strong enough so as to be unaffected by the energy lost in creating
polaritons.  This does not necessarily imply a 'high power' experiment, since
even low intensity beams have many photons, and the coupling between field and
polaritons is weak. The rate that new polaritons are created is $\epsilon_{+}$.

  The $({+})$ mode is dominated by its coherent driving, and so I write down
the equation for the evolution of its amplitude.  In contrast, the $({-})$ mode
is populated solely through the incoherent phonon scattering, and so has no 
coherent amplitude. As a result I use an equation for its intensity, 
which is non zero.  

 The complete equations for $\hat{p}_{+} = \hat{p}_{+} (0) $, $\hat{I}_{-} (k)$
can then be constructed.  Firstly, however, I can simplify them with the
physical considerations given above by removing the weakest decay terms and the
small thermal contributions.  The phonon scattering itself is very weak. 
Because of this, we also have $ \gamma_L (0) \gg \sum_k \chi (k) \left| d_{+}
d_{-} ^* (k) \right| ^2 $.  In the typical situation where $\hat{I}_{-} (k)$
remains small because the lower polariton is only populated by this scattering
process, this means that the effect of the phonon scattering terms in the
equation for $\hat{p}_{+}$ is negligible.

\begin{figure}
\includegraphics[width=80mm]{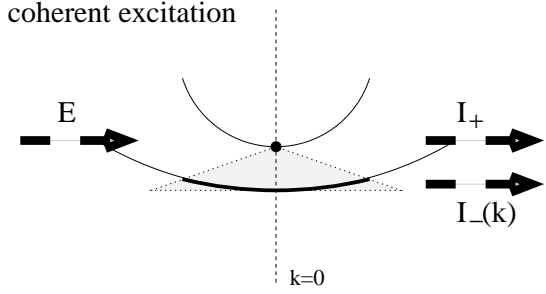}
\caption{ 
\label{Graph-CE}
{ 
The coherent excitation model.  The thin curves are the
polariton branches.  The triangular shaded area gives the allowed 
energies and in-plane momenta of the acoustic phonons.  The overlap 
of these two is the set of
possible final states given that the initial state was at $k=0$.  The left hand
side dashed arrow represents the coherent excitation $E = \left| \epsilon_{+}
\right| ^2$ of the upper $k=0$ state, and the two right hand side arows are the
light emitted by decay from the polaritons through the cavity mode.
}
}
\end{figure}

  Figure \ref{Graph-CE} shows a diagram of the model.  The QSDE's
for it are

\begin{eqnarray}
d \hat{p}_{+}
&\approx&
\left\{
   \epsilon_{+}
-  \frac{1}{2} \gamma_L c_{+}^* c_{+} \hat{p}_{+}
\right\} dt
+ 
   \sqrt{ \gamma_L } c_{+} ^* d \hat{B}_L (0)
\nonumber \\
d \hat{I}_{-} (k)
&\approx&
\left\{
-     \gamma_L (k) c_{-} ^* (k) c_{-} (k) \hat{I}_{-}
\right.
\nonumber \\
&&
\left.
+
   \chi (k) \left| d_{+} d_{-} ^* (k) \right| ^2
   \hat{p}_{+} ^\dagger \hat{p}_{+}  \left[ \hat{I}_{-} (k) + 1 \right] 
\right\} dt
\nonumber \\
& &
+  \sqrt{ \gamma_L(k) } c_{-} ^* (k) 
   \hat{p}_{-} ^\dagger (k) d \hat{B}_L (k) 
\nonumber \\
&&
+  \sqrt{ \gamma_L(k) } c_{-}    (k) 
   d \hat{B}_L (k) ^\dagger     \hat{p}_{+} (k)
\nonumber \\
& &
+
   \sqrt{ \chi(k) } d_{+} ^* d_{-} (k) 
   \hat{p}_{+} ^\dagger \hat{p}_{-} (k) 
   d \hat{B}_P (0)  
\nonumber \\
&&
+ 
   \sqrt{ \chi(k) } d_{+} d_{-} ^* (k) 
   d \hat{B}_P ^\dagger (0)
   \hat{p}_{+} \hat{p}_{-} ^\dagger (k)
.
\end{eqnarray}

  Since we do not need the quantum mechanical noise correlations, 
we can approximate them by taking the expectation values
and factorising the different intensity moments (see section IV.A).
Using  $I_{-} = \left< \hat{I}_{-} \right>$ and $\alpha_{+} =
\left< \hat{p}_{+} \right>$, I can rewrite the equations as complex 
amplitude and rate equations

\begin{eqnarray}
\frac{d}{dt} \alpha_{+}
&\approx&
   \epsilon_{+}
 - \frac{\gamma_L}{2}
   \left| c_{+} \right| ^2 \alpha_{+}
,
\nonumber \\
\frac{d}{dt} I_{-} (k)
&\approx&
\left[ 
-  \gamma_L (k) 
   \left| c_{-} (k) \right| ^2
    I_{-} (k)
\right.
\nonumber \\
&+&
\left. 
   \chi (k) 
   \left| d_{+} d_{-} (k) \right| ^2
   \left| \alpha_{+} \right| ^2
   \left[ I_{-} (k) + 1 \right]
\right]
.
\end{eqnarray}

The exciton modes have steady state values of

\begin{eqnarray}
\alpha_{+}
&=&
\frac
   { 2 \epsilon_{+} }
   { \gamma_L \left| c_{+} \right| ^2 }
,
\nonumber \\
I_{-} (k) 
&\approx& 
\frac
{ \chi (k) \left| d_{+} d_{-} (k) \right| ^2 \left| \alpha_{+} \right| ^2 }
{ \left| c_{-} (k) \right| ^2 \gamma_L (k) }
,
\end{eqnarray}

where the equation for $I_{-} (k)$ has been further simplified by reusing the
fact that the phonon scattering is a weak process, with $\left| c_{-} (k)
\right| ^2 \gamma_L (k) \gg \sum_k \chi (k) \left| d_{+} d_{-} (k) \right| ^2$
and small $\left| \alpha \right| ^2$.  Each polariton mode emits photons at a
rate given by the product of its decay rate $ \left| c_{\pm} \right| ^2\gamma_L
(k)$ and the occupation $\left| \alpha_{+} \right| ^2$ or $I_{-}$.  For the
upper $({+})$ mode this is just $\left| c_{+} (k) \right| ^2 \gamma_L \left|
\alpha_{+} \right| ^2$.  Using the $({+})$ mode population, the $({-})$ modes
have populations of

\begin{eqnarray}
I_{-} (k) 
&=& 
\frac
{ \chi (k) \left| d_{+} d_{-} (k) \right| ^2 \left| \epsilon_{+} \right| ^2 }
{ 
  \left| c_{+} \right| ^2 \gamma_L 
  \left| c_{-} (k) \right| ^2 \gamma_L (k) 
}
.
\end{eqnarray}

The ratio of cavity decay intensities between all $({-})$
modes and the $k=0$ $({+})$ mode is 

\begin{eqnarray}
R
&=& 
\sum_k
\frac
 { \chi (k) \left| d_{+} d_{-} (k) \right| ^2 }
 { \left| c_{+} \right| ^2 \gamma_L }
.
\end{eqnarray}

 From this expression we can see that the relative strengths of the upper and
lower exciton emission peaks is controlled largely by the strength of the
phonon scattering between the two exciton branches.  As a result, the upper
exciton peak is larger than the lower exciton peak.  Note that in the limits
considered here, this result is identical to that for the PL system.

\end{subsection}

\end{section}


\begin{section}{Stimulated scattering experiment}


  The stimulated scattering is perhaps the most interesting effect to be
predicted by this quantum model of phonon scattering.  Given this, how would it
be possible to measure its effects experimentally?  Since the size of the
effect is small due to the weak exciton-phonon coupling, it will be difficult
to separate the small depletion due to stimulated scattering (the signal) from
the optical decay from the polaritons (the background).

\begin{figure}
\includegraphics[width=80mm]{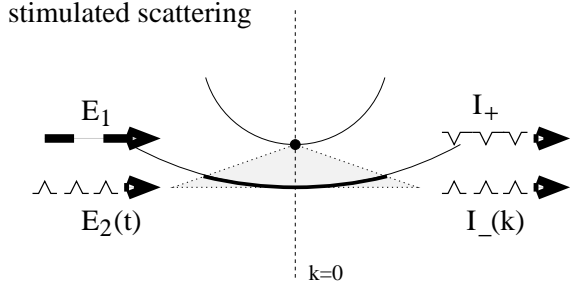}
\caption{ 
\label{Graph-SS}
{ 
The stimulated scattering experiment.  The thin curves are
the polariton branches.  The triangular shaded area gives the allowed 
energies and in-plane momenta of the acoustic phonons.  The overlap 
of these two is the set of
possible final states given that the initial state was at $k=0$.  The left hand
side dashed arrow represents the coherent excitation of the polariton states. 
The upper arrow is a continuous driving field $E_1 = \left| \epsilon_{+}
\right| ^2$ exciting the upper $k=0$ state, and the lower is the pulse train
that periodically excites the chosen lower state at $k=k_0$.  The two right
hand side arows are the light emitted by decay from the polaritons through the
cavity mode.  The upper level emission has a series of dips caused by the
stimulated scattering induced by the periodic excitation of the lower polariton
level.
}
}
\end{figure}

  Consider a simple polariton system with two branches.  There are cavity
losses, coherent driving fields tuned to the polariton energies, and
inter-branch coupling by acoustic phonon processes.  Recombination is neglected
as the cavity decay is generally much larger.  The upper branch is driven by a
CW coherent field at its $k=0$ energy.  This means that in the absence of any
thermal phonons, there are no phonon processes along the upper branch to
complicate matters.  For simplicity I also only include a single $k$ value on
the lower branch $k=k_0$, in an approximation which is justified at the end of
this section.  The lower level is driven periodically by very short laser
pulses that are well separated by a time interval $\tau$, with $\tau \gg
1/\gamma_{1}, 1/\gamma_{2}$.  Between each pulse, the system returns back to
its ``relaxed state'', which is that for an undriven lower level.  Note that
this relaxed state is the same as that worked out in the coherently pumped
bottleneck calculation in section VI.B. Figure \ref{Graph-SS} shows a diagram
of the model.  

Choosing a $k_0 \neq 0$ angle on the lower branch means we can separate the
output signal from the upper and lower branches by detection angle.  This is
important because the phonon effect is very small, and could be easily masked
by the addition of a strong background due to the driving fields.  Also, by
varying the lower branch $k_0$ that we select, we can probe how the the phonon
coupling strength varies with transverse wavevector $k$.

The coherent driving terms mean that is is most convenient to use the 
c-number amplitude equations.  The equations are ($N_P = 0$)

\begin{eqnarray}
\frac{d}{dt} \alpha_{+}
&=&
   \epsilon_{+}
 - \frac{ \gamma_{+} }{2} \alpha_{+}
 - \frac{ \chi' }{2} \alpha_{+} 
                     \left( \alpha_{-} ^* \alpha_{-}  + 1 \right)
\nonumber \\
&&
 + \imath \sqrt{ \chi' \alpha_{+} \alpha_{-} } W_{2}
,
\nonumber \\
\frac{d}{dt} \alpha_{-}
&=&
   \epsilon_{-} (t)
 - \frac{ \gamma_{-} }{2} \alpha_{-}
 + \frac{ \chi' }{2} \alpha_{-} \alpha_{+} ^* \alpha_{+}
\nonumber \\
&&
+ \sqrt{ \chi' \alpha_{+} ^* \alpha_{+} } W_{B}
+ \imath \sqrt{ \chi' \alpha_{+} \alpha_{-} } W_{2} ^*
\label{stim-alpha-eqns}
\end{eqnarray}

  The equations can be explicitly converted into those for a polariton system
by setting
$\alpha_{+} \rightarrow \alpha_{+} (0) $, 
$\alpha_{-} \rightarrow \alpha_{-} (k_0) $, 
$\gamma_{+} = \gamma_L c_{+} (0) ^* c_{+} (0) $, 
$\gamma_{-} = \gamma_L c_{-} (k_0) ^* c_{-} (k_0) $, 
$\epsilon_{+} = \epsilon_{+}  c_{+} (0) ^* $,
$\epsilon_{-} = \epsilon_{-}  c_{-} (k_0) ^* $,
$\chi' = \chi (0, -k_0) d_{+} (0) d_{-} (k_0) ^* $. 
It is also possible to do this kind of experiment in a two branch exciton system
by using angled and appropriately tuned laser beams for direct excitation of 
the desired point on the exciton branch.  Although in a PL setup a
steady source of excitation might be supplied by the combination of free
carriers into polaritons, this is neither $k$ specific, nor can it be rapidly
modulated, nor can the individual branches be selectively excited.

I now expand equations (\ref{stim-alpha-eqns}) about their relaxed state, and
write

\begin{eqnarray}
\alpha_{+} (t) &=& \alpha_{+0} + \beta_{+} (t)
,
\nonumber \\
\alpha_{-} (t) &=& \alpha_{-0} + \beta_{-} (t)
,
\label{stim-relax-deltas}
\end{eqnarray}

where $\alpha_{\pm 0}$ are the 'relaxed states' of the two levels.  The
depletion $\beta_{+}$ of the upper level can be treated as a small
perturbation, since it will be small compared to its level of excitation
$\alpha_{+0}$.  The relaxed state is like that for the bottleneck
calculations so it has $\alpha_{-0}=0$ because there is no coherent driving 
and the excitation due to phonon scattering is incoherent.  The intensity 
$
I_{-0} = \alpha_{-0} ^* \alpha_{-0} 
\approx \chi' \left| \alpha_{+0} \right| ^2 / \gamma_{+} 
$
would be non zero because of the $W_2$ noise terms.  Because the excitation of
the lower level is very small so the change in its excitation $\beta_{-}$
cannot be treated as a perturbation.  Inserting equations
(\ref{stim-relax-deltas}) into the equations above (\ref{stim-alpha-eqns})
gives us

\begin{eqnarray}
\frac{d}{dt} \beta_{+}
&=&
\frac{1}{2} \gamma_{+}
\left[
 - \beta_{+}
\right.
\nonumber \\
&&
\left.
 - \frac{\chi'}{\gamma_{+}} \beta_{+} 
   \left(
      \alpha_{-0} ^* \alpha_{-0} 
    + \beta_{-}  ^* \alpha_{-0} 
    + \alpha_{-0} ^* \beta_{-} 
    + \beta_{-}  ^* \beta_{-} 
   \right)
\right.
\nonumber \\
&&
\left.
 - \frac{\chi'}{\gamma_{+}} \alpha_{+0} 
   \left(
      \beta_{-}  ^* \alpha_{-0} 
    + \alpha_{-0} ^* \beta_{-} 
    + \beta_{-}  ^* \beta_{-} 
   \right)
\right]
,
\nonumber \\
\frac{d}{dt} \beta_{-}
&=&
\epsilon_{-} (t)
- 
\frac{1}{2} 
\gamma_{+}
\frac{ \gamma_{-}} { \gamma_{+} }
\left[
   1
   -
   \frac{\chi'}{\gamma_{+}} \frac{ \gamma_{+}} { \gamma_{-} }
   \left(
      \alpha_{+0} ^* \alpha_{+0} 
  \right.
\right.
\nonumber \\
&&
\left.
  \left.
    + \beta_{+}  ^* \alpha_{+0} 
    + \alpha_{+0} ^* \beta_{+} 
    + \beta_{+}  ^* \beta_{+} 
   \right)
\right]
\beta_{-}
.
\label{stim-beta-eqns}
\end{eqnarray}

To do the calculation rigorously I rescale the time by $\gamma_{+}$ so that
everything could be expanded in powers of the small parameter 
$ g = \chi' / \gamma_{+}$.  
I have not defined new scaled variables, but just collected the factor 
together in an appropriate way.

The truncation to first order in $ \chi' / \gamma_{+} $ is easy if $ \gamma_{-}
/ \gamma_{+} \sim 1$, and $\beta_{-} ^* \beta_{-} \sim 1$.  The
condition on $\beta_{-}$ still allows the lower level driving pulses
to be non perturbative, but ensures that the depletion they cause in the upper
level {\em is} perturbative.  Further, since the depletion depends on the
coupling, I assume $\beta_{+}$ is at most of order $\chi' /
\gamma_{+}$.  This assumption is self consistent with the resulting solutions.

Note that these conditions are needed for a simple theoretical solution for
this problem, but the character of the results for more general conditions is
unchanged.  Truncating the equations to first order in $\chi' / \gamma_{+}$
involves neglecting any terms including two of $\chi' / \gamma_{+}$,
$\beta_{+}$, $\alpha_{-0}$.  The equations then can be easily
simplified down to

\begin{eqnarray}
\frac{d}{dt} \beta_{+}
&=&
 - \frac{\gamma_1}{2} 
\left[
   \beta_{+}
 - \frac{ \chi' }{ \gamma_{+} } \alpha_{+0} \beta_{-}  ^* \beta_{-} 
\right]
,
\nonumber \\
\frac{d}{dt} \beta_{-}
&=&
\epsilon_{-} (t)
- 
\frac{\gamma_{-}'}{2} \beta_{-}
.
\label{stim-beta-simple}
\end{eqnarray}

  I have introduced an effective decay rate $\gamma_{-}' = \gamma_{-} - 2
\chi' \alpha_{+0} ^* \alpha_{+0}$.

Now I have set up the model, I can investigate the effect of the lower level
driving pulses.  Imagine one of the $\epsilon_{-}$ pulses arrives at $t=0$. 
For simplicity, I assume that the width of the pulse is much shorter that
$1/\gamma_{-}$, so it can be treated as a delta function.  Laplace transform
methods then quickly give the simple exponential solution for $\beta_{-}$ as

\begin{eqnarray}
\beta_{-} (t) = a_{-} \exp \left( - \gamma_{-}' t / 2 \right)
.
\label{stim-beta2-solution}
\end{eqnarray}

The equation for $\beta_{+}$ is now completely defined, and is

\begin{eqnarray}
\frac{d}{dt} \beta_{+}
&=&
 - \frac{\gamma_{+}}{2} \beta_{+}
 - \frac{\chi'}{2} \alpha_{+0} \left| a_{-} \right| ^2 
                   \exp ( - \gamma_{-}' t )
\label{stim-beta1-equation}
\end{eqnarray}

This can also be solved  by Laplace transform methods, to give the pair of 
solutions

\begin{eqnarray}
\beta_{+}
&=& 
\frac{ \chi' \alpha_{+0} \left| a_{-} \right| ^2 } { 2 \gamma }
\exp \left( - \gamma_{+} t / 2 \right)
\left[ 1 - \exp ( - \gamma t ) \right]
,
\nonumber \\
\gamma &=& 2 \gamma_{-}' - \gamma_{+} \neq 0
\nonumber \\
\beta_{+}
&=& 
\frac{\chi \alpha_{+0} \left| a_{-} \right| ^2 } {2}
t
\exp \left( - \gamma_{+} t / 2 \right) 
,
\nonumber \\
\gamma &=& 2 \gamma_{-}' - \gamma_{+} = 0
.
\label{stim-beta1-solution}
\end{eqnarray}

\begin{figure}
\includegraphics[width=80mm]{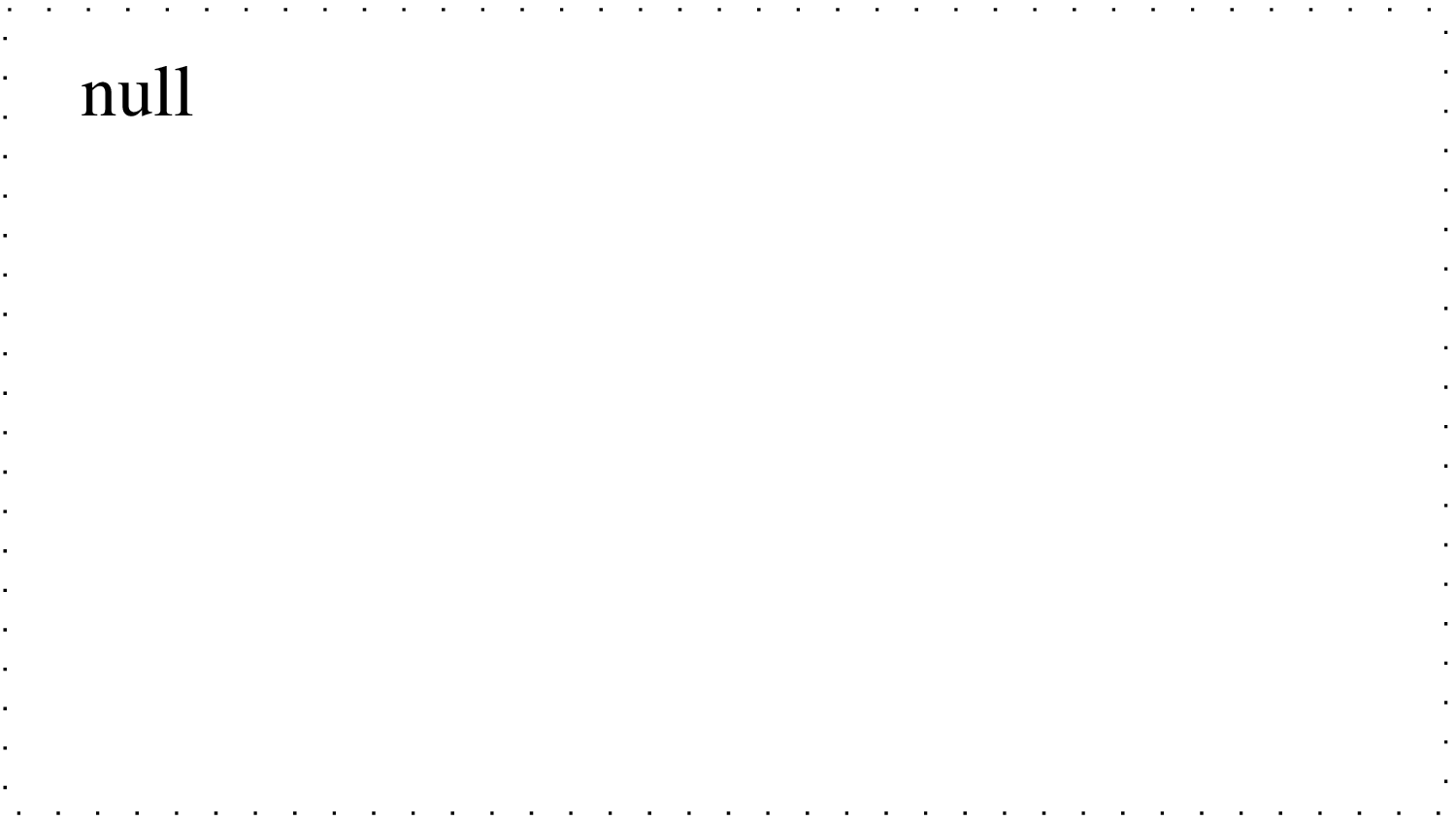}
\caption{
\label{Graph-SS-noiseless}
{
Graphs of (a) $\alpha-{+}$ and (b) $\alpha_{-}$ as functions of time for the 
model without the quantum noise terms.  The spectra for $\alpha-{+}$ (c)  
show the sidebands caused by the stimulated scattering.
}
}
\end{figure}

\begin{figure}
\includegraphics[width=80mm]{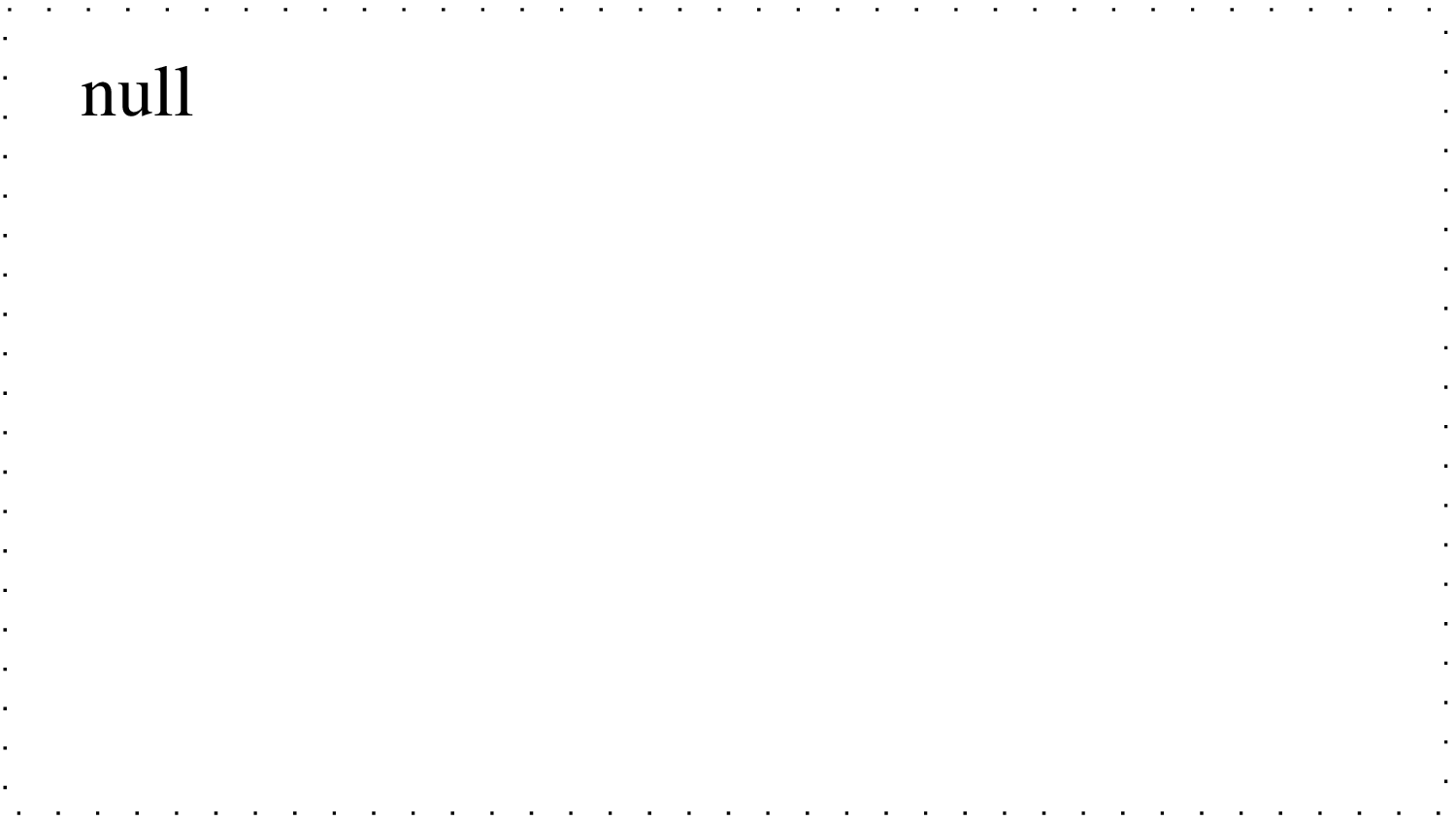}
\caption{
\label{Graph-SS-noisy}
{
Graphs of (a) $\alpha-{+}$ and (b) $\alpha_{-}$ as functions of time for the 
model with the quantum noise terms. The spectra for $\alpha-{+}$ (c)  
show the sidebands caused by the stimulated scattering.
}
}
\end{figure}

  So the result is a dip in the excitation of the upper level, with a duration
of about $1/\gamma_{+}$ and a relative depth proportional to $\chi' \left|
a_{-} \right| ^2$ -- note that this is consistent with the assumptions that
$\beta_{+}$ was of order $ \chi' / \gamma_{+} $ which was used in
the truncation to first order.  The size of the dips is proportional to the
pulse intensity (and not its amplitude).  A train of these input pulses
exciting the lower level will stimulate phonon scattering in bursts and produce
a corresponding series of dips in the upper level amplitude.  These dips will
show up in the optical decay from the upper mode, and could be detected by a
peak in the output spectrum at a frequency given by their repetition rate.  A
single dip would be hard to detect, because it would be easily masked by noise.
These dips show up in simulations of the equations -- figures 
\ref{Graph-SS-noiseless} and \ref{Graph-SS-noisy} 
show the behaviour of the outputs as a function of time for the both 
the simple noiseless case as well as the case with quantum noise included.

  Lastly, it is necessary to consider the effect of the other phonon scattering
processes in a real system, since up to now we have restricted it to a simple
two mode model, when in reality there will be scattering to other parts of the
lower branch as well, and so the excitation in the chosen lower level $k$ value
will depend on re-scattered excitation as well as the driving pulses and the
scattering from the upper level.  However, these processes will merely provide
a nearly constant background -- mostly due to the excitation cascading down
along the lower branch through $k=k_0$ down to $k=0$.  The only way they will
change the signal because of the pulse train is through second order processes.
The effect of the reduced excitation has to scatter down into the lower level
(one factor of $\chi'$), then back up (another factor of $\chi'$).  These types
of processes already occur in the simple model, but were neglected due to their
small size, and so will have negligible effect on the measurements.

Light could also scatter from the pulse train directly into the $k=0$ detector.
The strength of this could be measured by turning off the CW upper level
driving field and retaining the pulse train.  Also, any instability in the
driving laser at a frequency about that of the pulse repetition should be
reduced as much as possible to avoid contaminating the depletion signal.  This 
should be minimised by stabilising the laser, or by adjusting the repetition 
rate to fall in a quiet part or the laser spectrum.

\end{section}

\begin{section}{Conclusion}


 This paper has presented a derivation of equations for phonon processes using
the technique of quantum stochastic differential equations (QSDE's)
\cite{Gardiner-qn}.  The procedure had the advantage of being a systematic
derivation in which the approximations are clearly stated and whose effects are
generally well understood.  The technique was used to treat the phonon
interaction as a coupling between the two polariton branches and an infinite
reservoir of phonon modes, and properly accounted for the partly photonic
nature of the polariton.  The basic theoretical models of experimental
situations used here in conjunction with the weak nature of acoustic phonon
scattering predicts a ``bottleneck'' effect like that seen in recent low power
polariton experiments \cite{Tribe-etal}.  In addition, the rigourous quantum
approach to the phonon scattering interaction revealed some new ``stimulated
scattering'' terms whose existance would not have been obvious from a more
phenomenonological approach.

Further experimental investigations of this effect might take one of two paths.
Firstly, the detuning of exciton and photon might be varied to check the change
in strength of the bottleneck as the fractions of exciton and photon in the
polaritons vary. The change in strength should mean that the ratio of the
intensity peak areas should change in a corresponding way.  The second path,
and the one with more potential is to explore the use of coherent excitation.
With appropriate tuning of the driving field and perhaps even control of its
angle to the perpendicular, specific parts of the polariton branch could be
excited and the resulting angular dependence of the luminescence measured.  

An experiment was proposed to try to observe the ``stimulated scattering''
predicted by the quantum model used in this paper.  However, whether it is
possible to separate this effect out from the ``background'' of other processes
is debatable.  Theoretical avenues would involve treating the intra-branch
phonon scattering properly, and doing more detailed modelling of particular
experiments.  


\end{section}




\end{document}